\documentclass[final,3p,times,twocolumn]{elsarticle}
\usepackage{lineno}
\usepackage{graphicx}
\usepackage{subfigure}
\usepackage{amsmath}
\usepackage{color}
\usepackage{enumitem}
\usepackage{multirow}

\journal{Nuclear Instruments and Methods}

\begin{document}
\begin{frontmatter}

\title{Over saturation behavior of SiPMs at high photon exposure}

\author[smi]{L. Gruber}
\author[smi]{S. E. Brunner}
\author[smi]{J. Marton}
\author[smi]{K. Suzuki}
\address[smi]{Stefan Meyer Institute for Subatomic Physics, Austrian Academy of Sciences, Vienna, Austria}

\begin{abstract}

Several types of Silicon Photomultipliers were exposed to short pulsed laser light ($\sim$\,30\,ps FWHM) with its intensity varying from single photon to well above the number of microcells of the device. We observed a significant deviation of the output of SiPMs from the expected behavior although such response curve is considered to be rather trivial. We also noticed that the output exceeds the maximum expected pulse height, which should be defined as the total number of pixels times the single photon pulse height. At the highest light intensity ($\sim$\,500 times the number of pixels) that we tested, the signal output reached up to twice the maximum theoretical pulse height, and still did not fully saturate.

\end{abstract}

\begin{keyword}
Semiconductor photo detector
\sep
Silicon Photomultiplier (SiPM)
\sep
Multi Pixel Photon Counter (MPPC)
\sep
dynamic range

\end{keyword}

\end{frontmatter}

\section{Introduction}
\label{sec:introduction}
The Silicon Photomultiplier (SiPM) is a semiconductor photo detector which consists of multiple pixels (typically a few 100) of Avalanche Photodiodes working in Geiger-mode. Because of its characteristics such as compact size, low cost, insensitivity to magnetic fields, high photon detection efficiency (PDE) and high gain, the SiPM can be used in many different fields ranging from astrophysics, particle physics to medical imaging, as an alternative to vacuum Photomultiplier Tubes. 

Due to its design, the SiPM dynamic range should be limited to an order of the total number of pixels. This effect is reflected in a saturation behavior of the SiPM response. The relation between the number of incident photons on the detector surface ($N_{\it photon}$) and the number of fired pixels ($N_{\it fired}$) can be described by the following model:

\begin{equation}
  \centering
  N_{\it fired} = N_{\it total} \times \left[1 - \exp\left(-\frac{N_{\it photon}\times\mathrm{PDE}}{N_{\it total}}\right)\right]
  \label{eq:dynamicrange}
\end{equation}

with $N_{\it total}$, the total number of pixels of the SiPM. With increasing $N_{\it photon}$, the SiPM response curve, i.e. the relation between light input and SiPM output ($N_{\it fired}$), deviates from linearity, dependent on the PDE, and saturates at $N_{\it fired} = N_{\it total}$. Eq.~\ref{eq:dynamicrange} is valid for an ideal photosensor and an infinitely short light pulse. In a real SiPM, however, the response to incident light is influenced by several effects, such as after-pulsing, cross-talk, dark-noise and the pixel recovery. Therefore, the SiPM output is expected to deviate from the response curve as given by Eq.~\ref{eq:dynamicrange}. 

As presented in the following sections, we came across to observe a deviation between the SiPM output and the expected response, which cannot be explained only by the above effects. We measured the response curve for various SiPMs, all with 1\,mm$^2$ sensitive area but different number of pixels and from different vendors. The models tested are the Hamamatsu MPPC S10362-11-100U with 100 pixels and S10362-11-050U with 400 pixels, the SSPM-0611B1MM-TO18 from Photonique\footnote{Photonique SA has suspended its operations. Product information can be found here: http://www.photonique.ch/LEGACY.} with 556 pixels and a Zecotek MAPD-1 with 560 pixels. The main parameters of all tested devices are summarized in Table \ref{tab:SiPMparameters}.

\begin{table*}[t]
  \centering
  \begin{tabular}{ l  c  c  c  c }
  \multicolumn{1}{c}{\multirow{2}{*}{Parameter}} & \multicolumn{2}{c}{Hamamatsu MPPC} & Photonique SSPM & Zecotek MAPD \\
			     & S10362-11-100U & S10362-11-050U & 0611B1MM-TO18 & MAPD-1 \\ \hline \hline
  Active area [mm$^{2}$] & 1$\times$1 & 1$\times$1 & 1$\times$1 & 1$\times$1 \\ 
  Number of pixels & 100 & 400 & 556 & 560 \\ 
  Pixel size [$\mu$m$^{2}$] & 100 $\times$ 100 & 50 $\times$ 50 & -- & -- \\ 
  Fill factor [$\%$] & 78.5 & 61.5 & $>$ 70 & -- \\ 
  PDE [$\%$ @ $400\,\mathrm{nm}$] & 72 & 47 & 18 & 15 \\ 
  Capacitance [pF] & 35 & 35 & 40 & 75.6 \\ 
  Breakdown voltage [V] & 69.45 & 68.65 & 27.80 & 34.00 \\ 
  Operating voltage $V_{\it bias}$ [V] & 69.95 & 69.85 & 29.00 & 34.70 \\ 
  Gain @ $V_{\it bias}$ & 1.1 $\times$ 10$^{6}$ & 6.6 $\times$ 10$^{5}$ & 5.4 $\times$ 10$^{5}$ & 5.9 $\times$ 10$^{5}$ \\ \hline
  \end{tabular}
  \caption{Main SiPM parameters. The breakdown voltage, $V_{bd}$, has been measured. The operating voltage $V_{\it bias}$ is typically set $\sim$\,1\,V above $V_{bd}$. The exact values and the corresponding gain are given. Other parameters are taken from the data sheets~\cite{hamamatsu, photonique, orth}. The PDE given by Hamamatsu includes effects from cross-talk and after-pulsing. There are several other measurements of the PDE available, but the results are also known to strongly depend on the operating conditions, e.g. over-voltage and temperature, and the measurement procedure. Therefore, we refer to the values given by the companies.}
  \label{tab:SiPMparameters}
\end{table*}

\section{Setup and Method}
\label{sec:method}
\begin{figure}[t]
  \centering
  \includegraphics[width=0.47\textwidth]{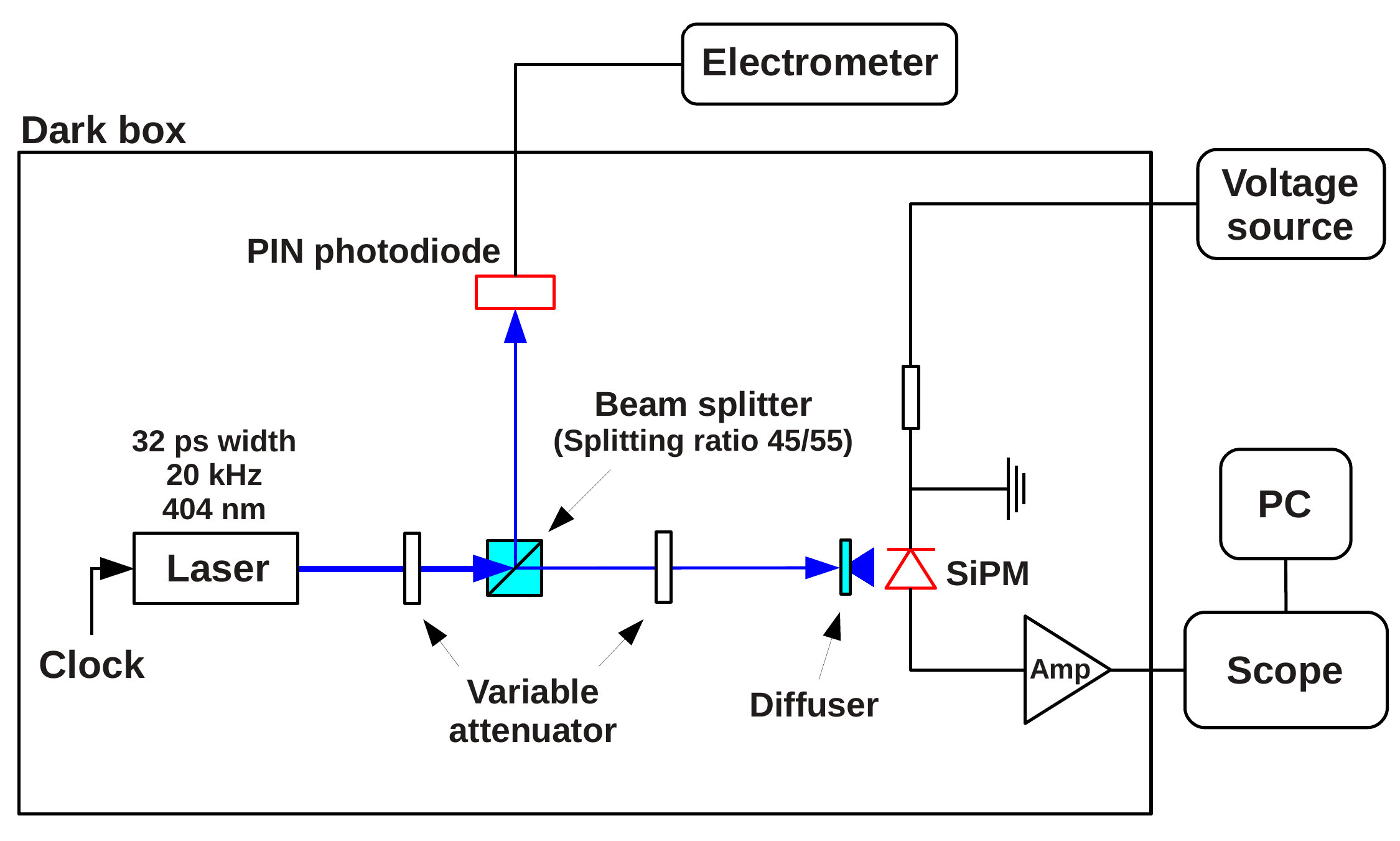}
  \caption{Schematic of the measurement setup.}
  \label{fig:setup}
\end{figure}

To measure the response curve, the SiPMs were exposed to short light pulses with intensities ranging from single photon up to several ten thousand. The measurement setup is shown schematically in Fig.~\ref{fig:setup}. All tests were done at room temperature ($\sim$\,25\,$^{\circ}$C). As light source we used a pulsed laser with 32\,ps pulse width (FWHM) from Advanced Laser Diode Systems. The emission wavelength of the laser head (PIL040) is $\lambda=404$\,nm. The repetition frequency was set to a level of 20\,kHz, to have a time interval between two laser pulses well above the SiPM cell recovery time. After passing a variable optical attenuator, the laser pulses were split using a beam splitter with a splitting ratio of 45:55 (45\,\% reflectivity, 55\,\% transmission). One path of the beam is targeted at a Hamamatsu S5971 PIN photodiode for monitoring the light intensity. The current of the PIN photodiode was measured using a Keithley 6517 electrometer. After passing another variable optical attenuator, the second beam was directed to a diffuser in order to homogeneously distribute the light on the SiPM active area. The second attenuator in between beam splitter and SiPM is needed to deal with the different sensitivities of the SiPM and the photodiode.

The SiPM signal was amplified by using a Photonique AMP-0611 preamplifier~\cite{photonique}. The DC voltage supply was set to 5\,V. The linearity of the preamplifier was confirmed by measuring the preamplifier response to defined input pulses. Within the whole input range we tested, a linear behavior of both, the output pulse height as well as the output charge, was found. The measurement resulted in a gain of about 23. The operating voltage of the sensor, $V_{\it bias}$, was typically set to $V_{\it over} \sim 1$\,V above the breakdown voltage, $V_{\it bd}$, which had been determined in a separate measurement. The values are given in Table~\ref{tab:SiPMparameters}. The corresponding gain of the SiPM, $G$, can be estimated by $G = C_{\it pix} \cdot V_{\it over}/q_{\it e} = C_{\it pix} \cdot (V_{\it bias} - V_{\it bd})/q_{\it e}$, with $C_{\it pix}$ being the pixel capacitance and $q_{\it e}$ the elementary charge. The operating voltages given in Table~\ref{tab:SiPMparameters} were selected in order to operate the SiPMs at low to moderate gain and therefore low noise (dark-noise, after-pulsing, cross-talk).

\begin{figure}[t]
  \centering
  \includegraphics[width=0.4\textwidth]{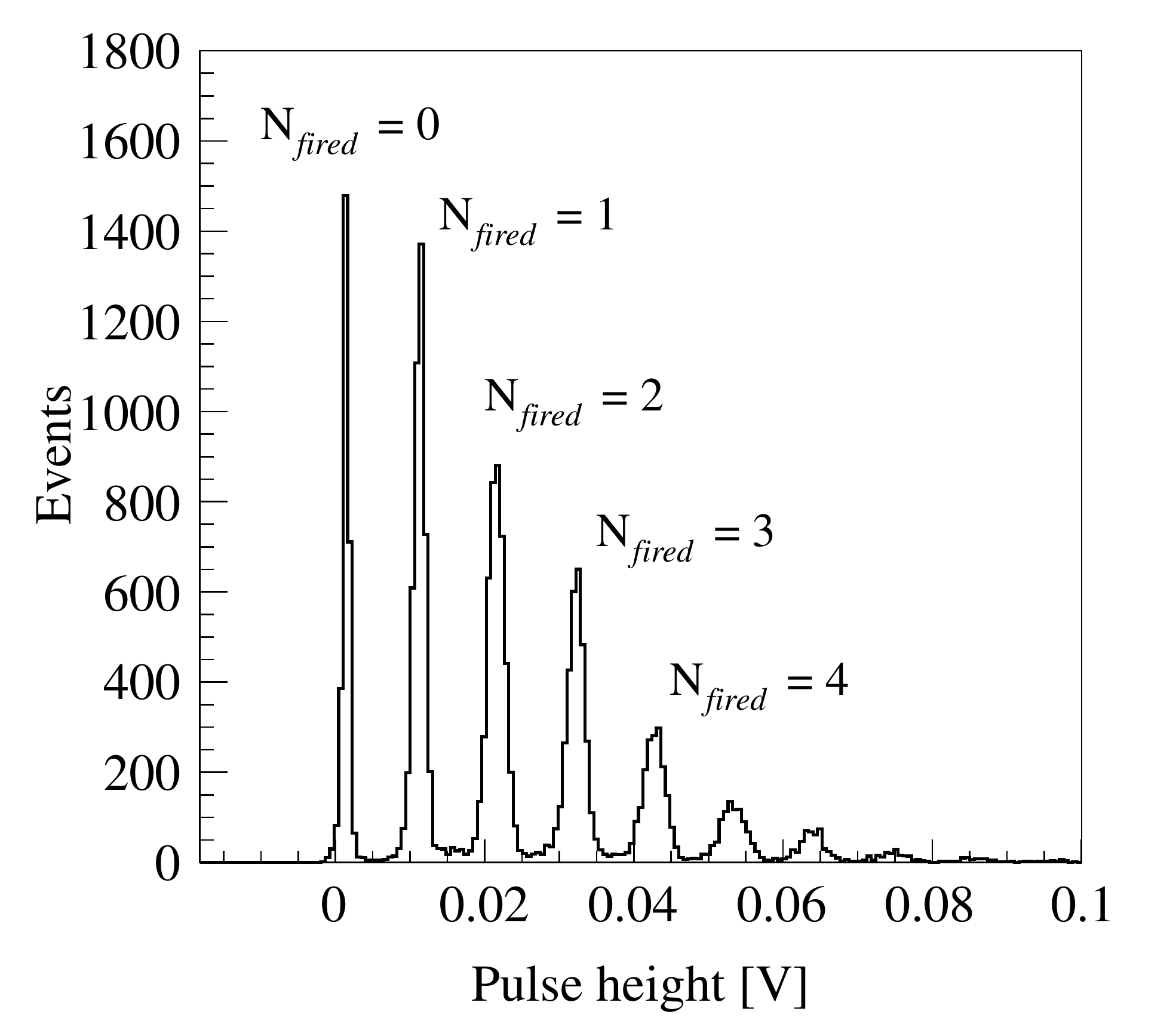}
  \caption{Single photon spectrum of the Hamamatsu MPPC with 100 pixels operated at $V_{over} = 1\,\mathrm{V}$. The peaks correspond to a certain number of fired pixels.}
  \label{fig:singlephotonspectrum}
\end{figure}

The SiPM response, i.e. the number of fired pixels, $N_{\it fired}$, was determined by measuring the average output pulse height with the LeCroy WavePro 735Zi digital oscilloscope. In order to estimate $N_{\it fired}$ from the measured pulse height, the output signal of a single fired pixel must be determined. This is done at low light intensity by filling the pulse height values into a histogram, as shown in Fig.~\ref{fig:singlephotonspectrum}. Each peak corresponds to a certain number of fired pixels ($N_{\it fired}$). By fitting the spectrum and extracting the distance between adjacent peaks, the pulse height corresponding to a single fired pixel can be determined several times. 

\begin{figure}[t]
  \centering
  \includegraphics[width=0.30\textwidth]{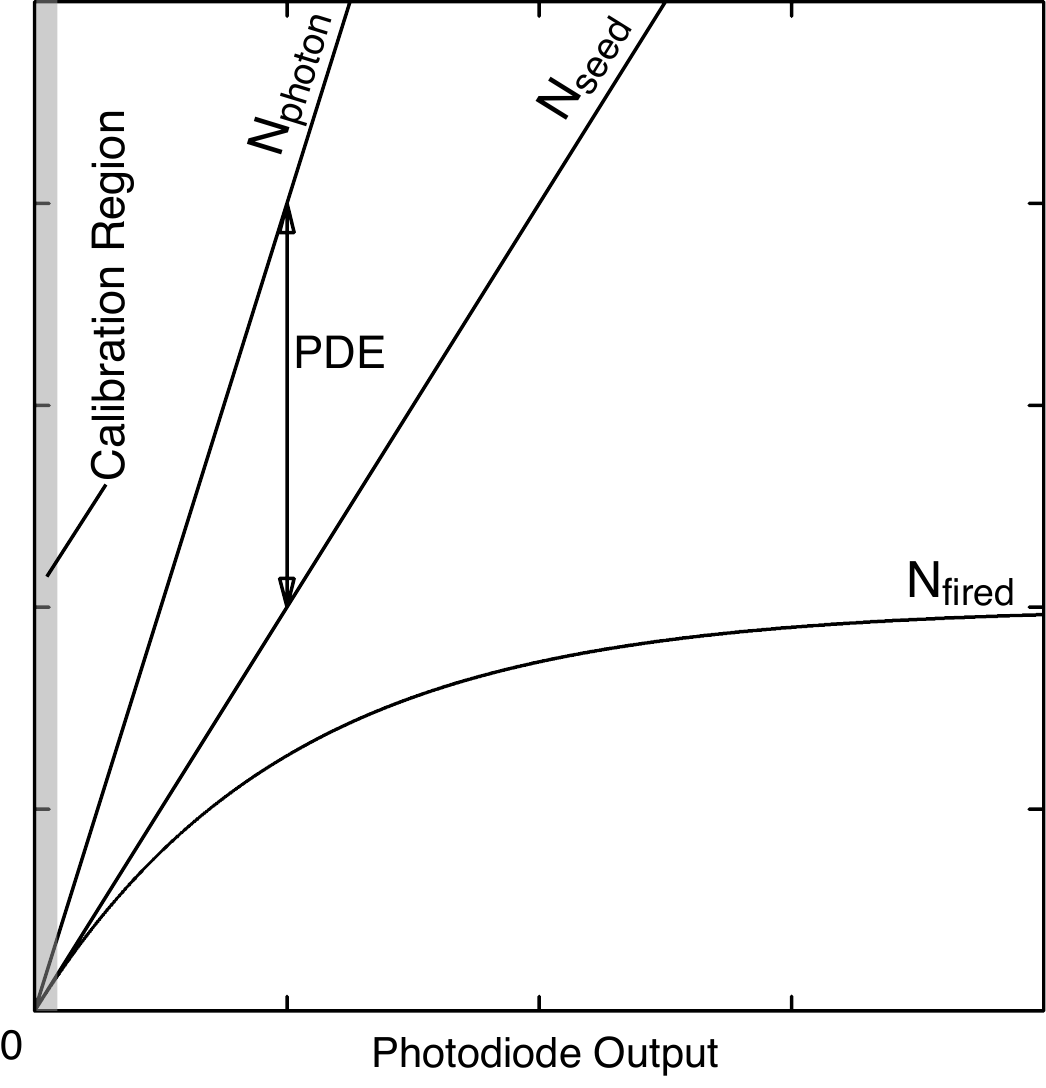}
  \caption{Schematic of the calibration method.}
  \label{fig:method}
\end{figure}

The PIN photodiode was calibrated at very low light intensities ($N_{\it fired} < 10$, in case of Hamamatsu 100U, $N_{\it fired} < 20$, for the others), where one can expect a linear behavior of the response, due to the homogeneous distribution of input photons on the sensor surface. The calibration procedure is illustrated schematically in Fig.~\ref{fig:method}. For interpretation of the data, we introduce the average number of "seeds", $N_{\it seed}$, which is the average number of photons arriving at the sensitive area of the SiPM, that could trigger an avalanche unless the cells had been fired already. The number of fired pixels, $N_{\it fired}$, is the main observable of the measurement. $N_{\it fired}$ can be determined by measuring the signal pulse height, as described before. In the calibration region $N_{\it seed} = N_{\it fired}$, thus $N_{\it seed}$ and the linear relation between the photodiode output current and the number of "seeds" can be determined and in the following extrapolated to higher light intensities. The relation between $N_{\it seed}$ and the number of incident photons, $N_{\it photon}$, is given by $N_{\it photon} = N_{\it seed}/\mathrm{PDE}$. In order to avoid the use of a PDE, which depends on the temperature, the operation voltage and the way it is measured, we plot $N_{\it fired}$ as a function of $N_{\it seed}$ and compare different types of SiPMs. This is a more "natural" representation, because only measured quantities are involved and therefore the actual PDE of the sensor is included a priori. The expectation curve then appears as Eq.~\ref{eq:dynamicrange_noPDE}:

\begin{equation}
  \centering
  N_{\it fired} = N_{\it total} \times \left[1 - \exp\left(-\frac{N_{\it seed}}{N_{\it total}}\right)\right]
  \label{eq:dynamicrange_noPDE}
\end{equation}

As mentioned before, we are determining $N_{\it fired}$ by measuring the signal pulse height rather than the output charge. The idea behind is to minimize the influence of delayed correlated noise, like after-pulsing and cross-talk related to this after-pulsing, as well as the influence from dark-noise. In course of the measurements we also monitored the integrated charge and checked that there's no apparent change in the signal shape for all light intensities that were tested. Since we are operating the SiPM at low gain and therefore low noise, we realized that the measured SiPM response is hardly affected by the method used to determine $N_{\it fired}$ (pulse height or charge measurement). The impact of the diverse noise effects on the measurement results is estimated in the following.

\begin{description}[leftmargin=0cm]
  \item[Dark-noise:] The pulse height measurement is not influenced by dark-noise, unless the dark-counts occur within the rise time of the signal, which is about 2\,-\,3\,ns. Since the dark-count rate is typically below 1\,MHz for a 1\,mm$^2$ SiPM, the probability for a dark-count to occur within this time interval is in the order of a few per mill and therefore the impact on the measurement is negligible.
  \item[After-pulsing:] The measured output pulse height could be affected by after-pulses that happen within a very short time interval after the initial breakdown, within the signal rise time, however, the influence is small due to the finite recovery time of the pixels ($\sim$\,200\,ns till full recovery for the Hamamatsu 100U, $\sim$\,50\,ns for the Hamamatsu 050U~\cite{handbook, oide}) and a certain after-pulse probability. Taking the recovery process into account, an after-pulse happening 2\,-\,3\,ns after the initial avalanche in a microcell may give rise to an additional output of $\sim$\,5\,-\,15\,\% of the single photon signal, depending on the pixel size. Assuming additionally an after-pulse probability of 10\,\% for the MPPCs with 100\,$\mu$m$^{2}$ and 50\,$\mu$m$^{2}$ pixel size~\cite{schultzcoulon}, the total output signal of the SiPM might be overestimated finally by $\sim$\,1\,-\,2\,\% due to fast after-pulsing.
  \item[Cross-talk:] The cross-talk probability is known to be $\sim$\,5\,\% at the operating conditions we are using~\cite{schultzcoulon}. The impact of cross-talk related to fast after-pulsing is negligible because of the relatively small effect of after-pulsing itself. In fact, the measurement is mainly influenced by almost instantaneous cross-talk following the initial photon input (laser pulse). However, this effect is suppressed especially at high light intensities (in the non-linear range) because the pixel occupancy is already nearly 100\,\%. In this range most of the measurements were performed. In the calibration region the effect is also small. The SiPM response ($N_{\it fired}$) might be overestimated by $\sim$\,5\,\% in the linear region due to cross-talk and thus also $N_{\it seed}$, since we use $N_{\it seed} = N_{\it fired}$ to determine the number of "seeds". Another uncertainty in the photon count ($N_{\it seed}$) comes from statistical fluctuations in the number of photons detected.
\end{description}

Because of the above considerations, the uncertainty of the SiPM output ($N_{\it fired}$) due to after-pulsing, cross-talk and dark-noise is $<$\,2\,\% for most light intensities that were used in the measurements, i.e. beyond the linear region of the response. In this regime, the response is mainly influenced by after-pulsing. For low light intensities, also cross-talk starts to play a role and $N_{\it fired}$ and $N_{\it seed}$, respectively, may be overestimated by $\sim$\,5\,\%. There should be no impact of the recovery process on the measured SiPM response, since the incident light pulse duration is rather short ($\sim$\,30\,ps) compared to the pixel recovery time (few 10\,ns). Therefore, one can use Eq.~\ref{eq:dynamicrange} and Eq.~\ref{eq:dynamicrange_noPDE}, respectively, to describe the measured response curves. 

\begin{figure}[t]
  \centering
  \includegraphics[width=0.4\textwidth]{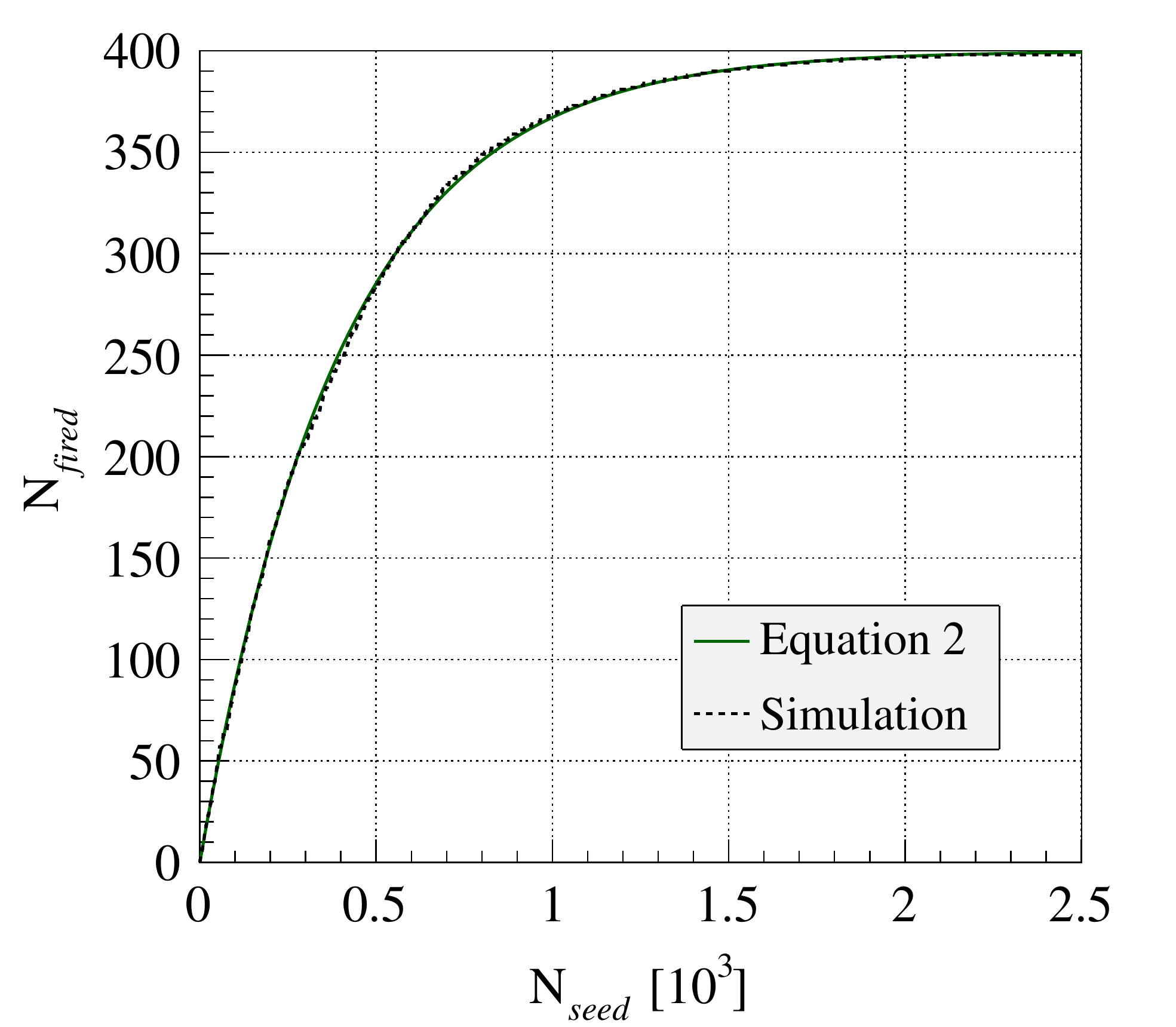}
  \caption{Comparison between simulated and modeled response of an SiPM with $N_{\it total} = 400$.}
  \label{fig:simulation}
\end{figure}

For verification of the model equation, we performed a Monte Carlo simulation including a geometrical consideration and obtained a response curve. The result is shown in Fig.~\ref{fig:simulation} together with the curve expected from the model, as given by Eq.~\ref{eq:dynamicrange_noPDE}. The two curves are found essentially identical, indicating that the approximation of the SiPM response using the model equation is reasonable.

\section{Results}
\label{sec:results}
\begin{figure*}[t]
  \centering
  \includegraphics[width=0.34\textwidth]{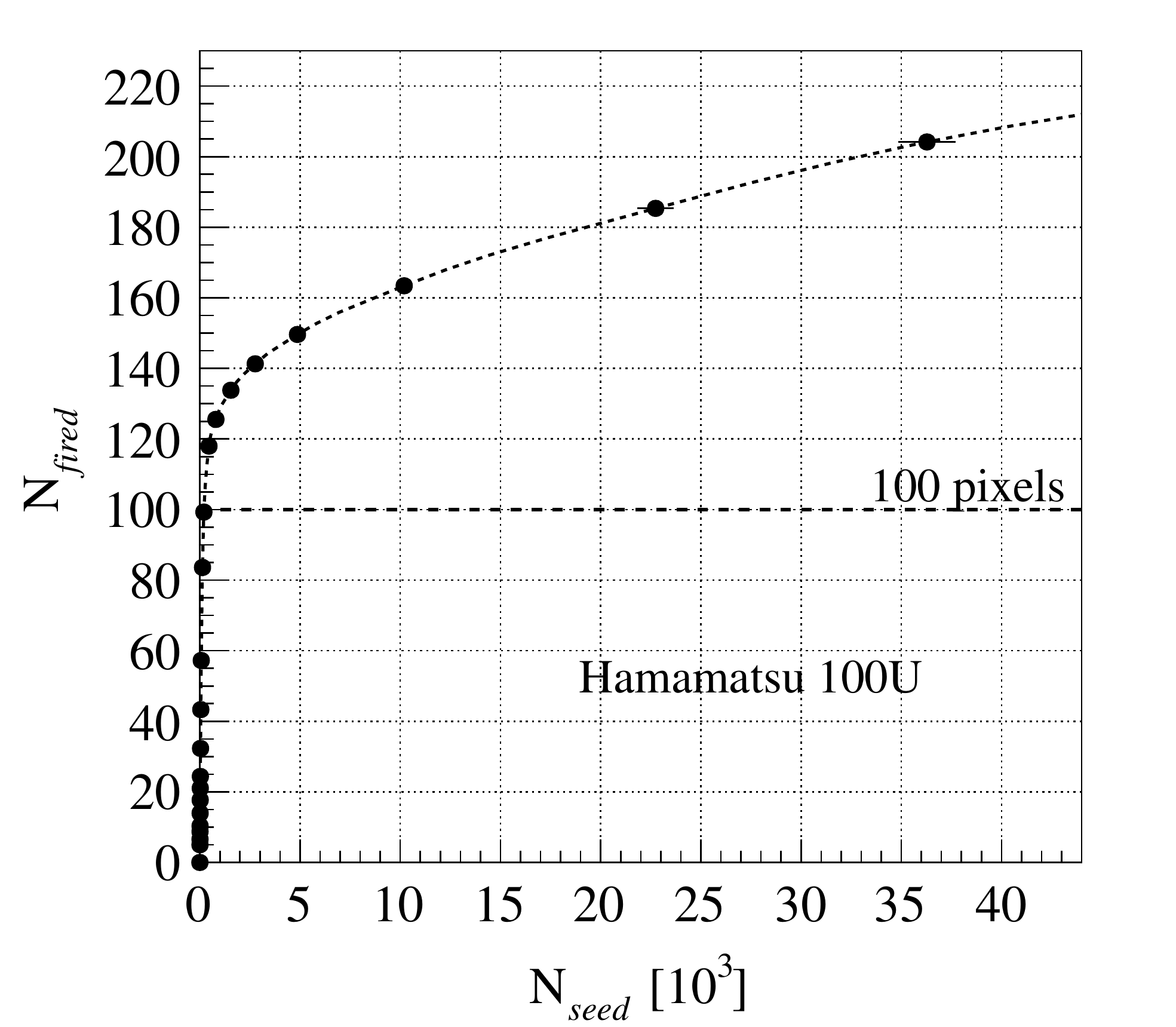}
  \includegraphics[width=0.34\textwidth]{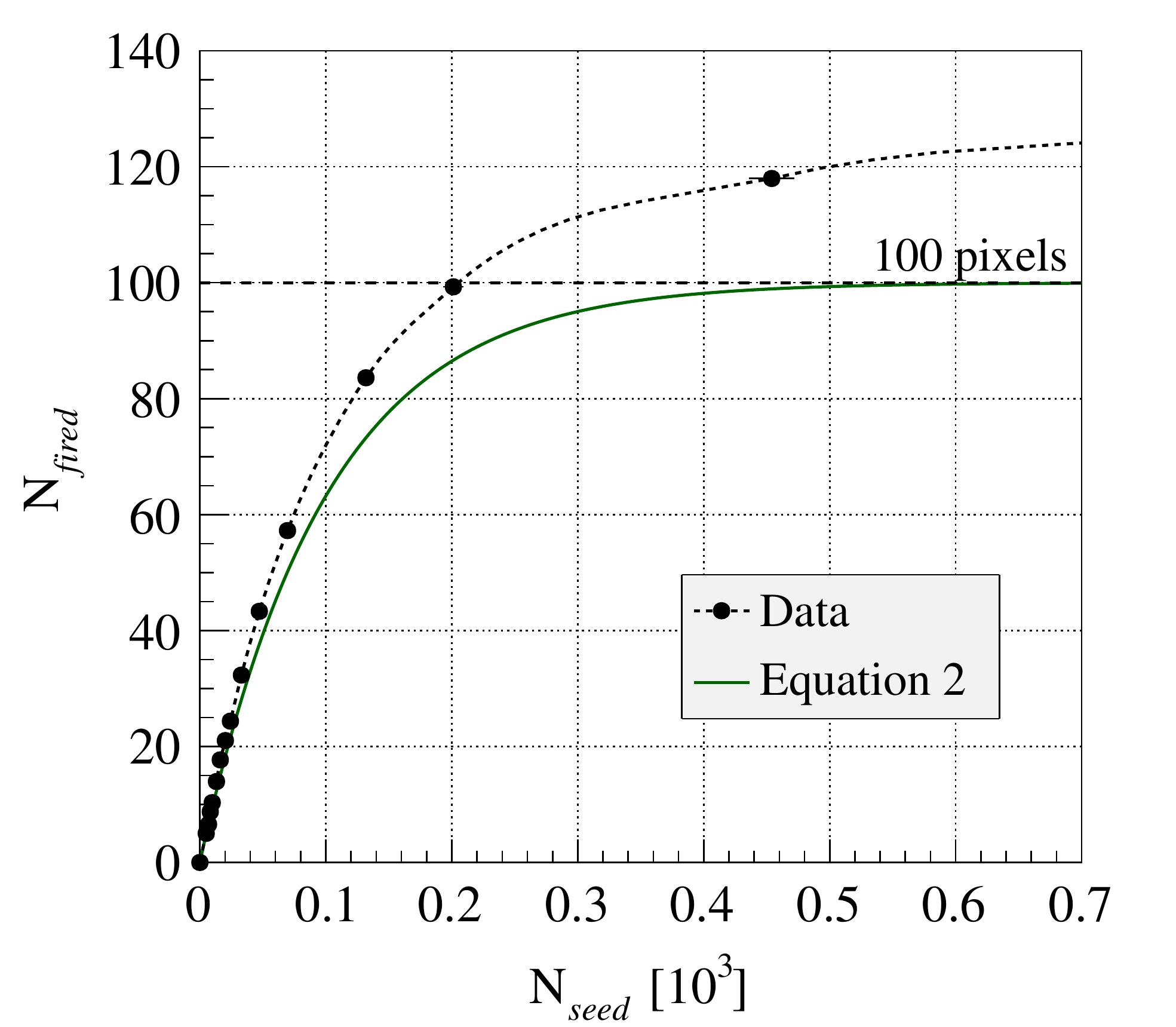}
  \includegraphics[width=0.34\textwidth]{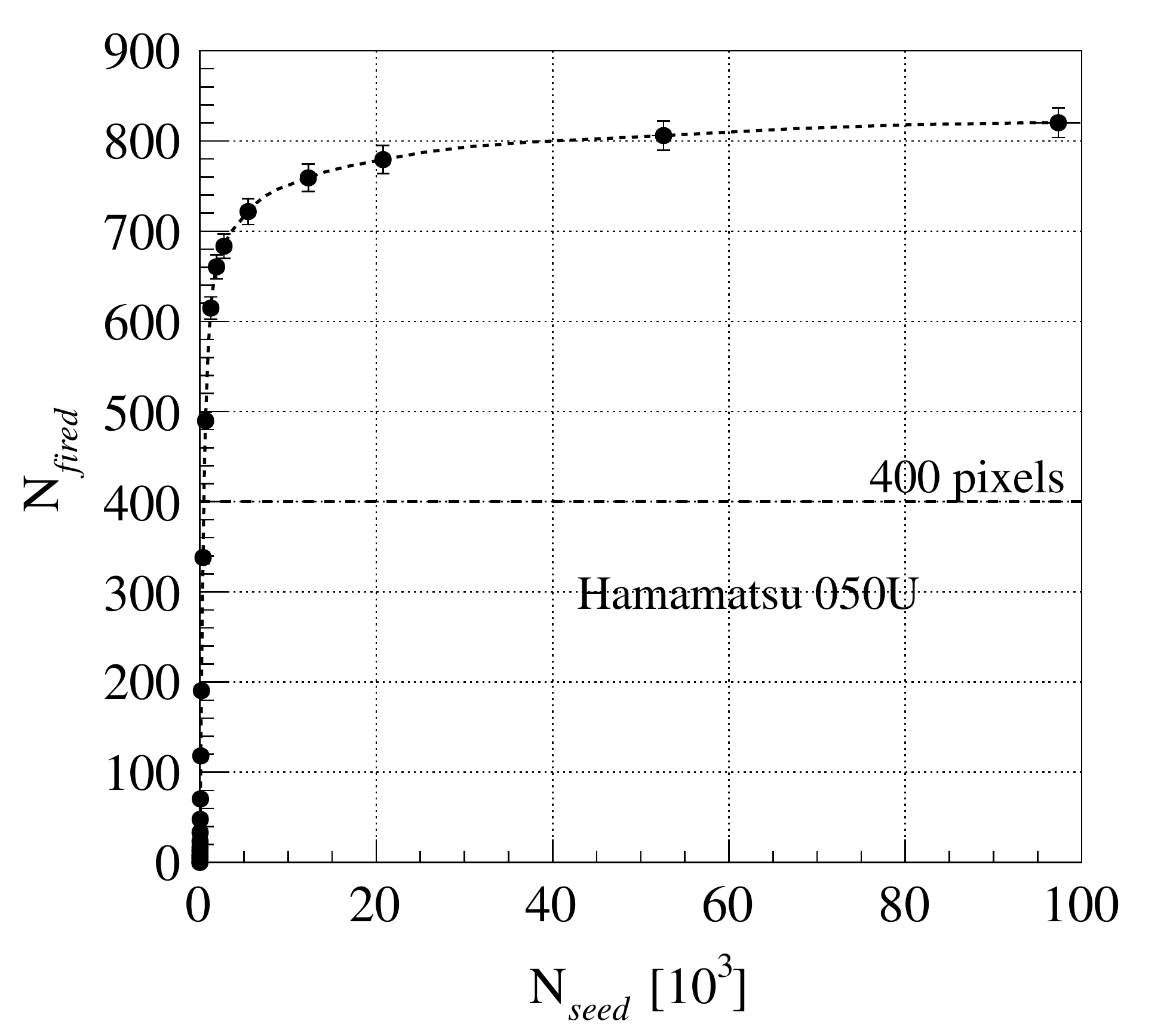}  
  \includegraphics[width=0.34\textwidth]{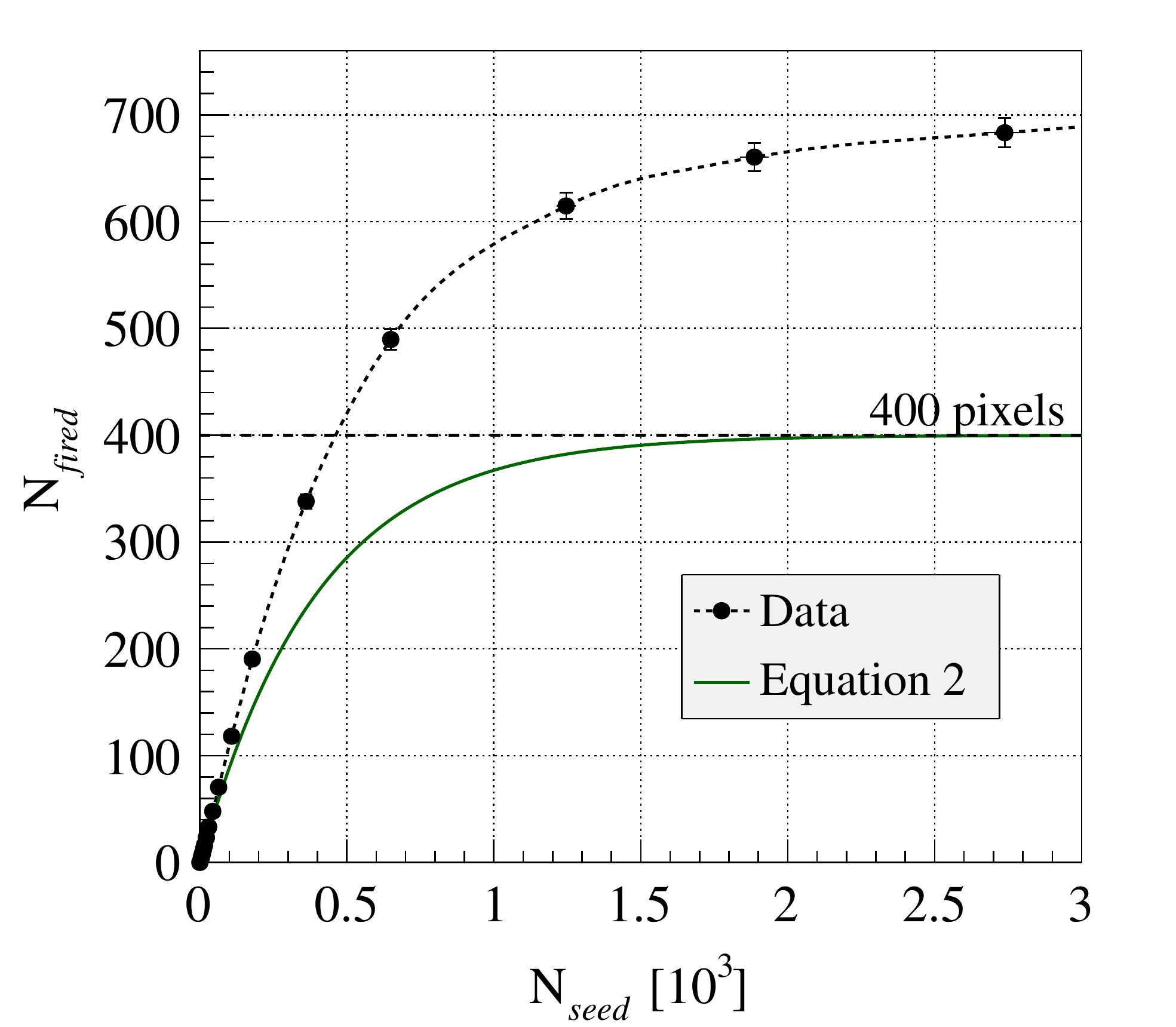} 
  \includegraphics[width=0.34\textwidth]{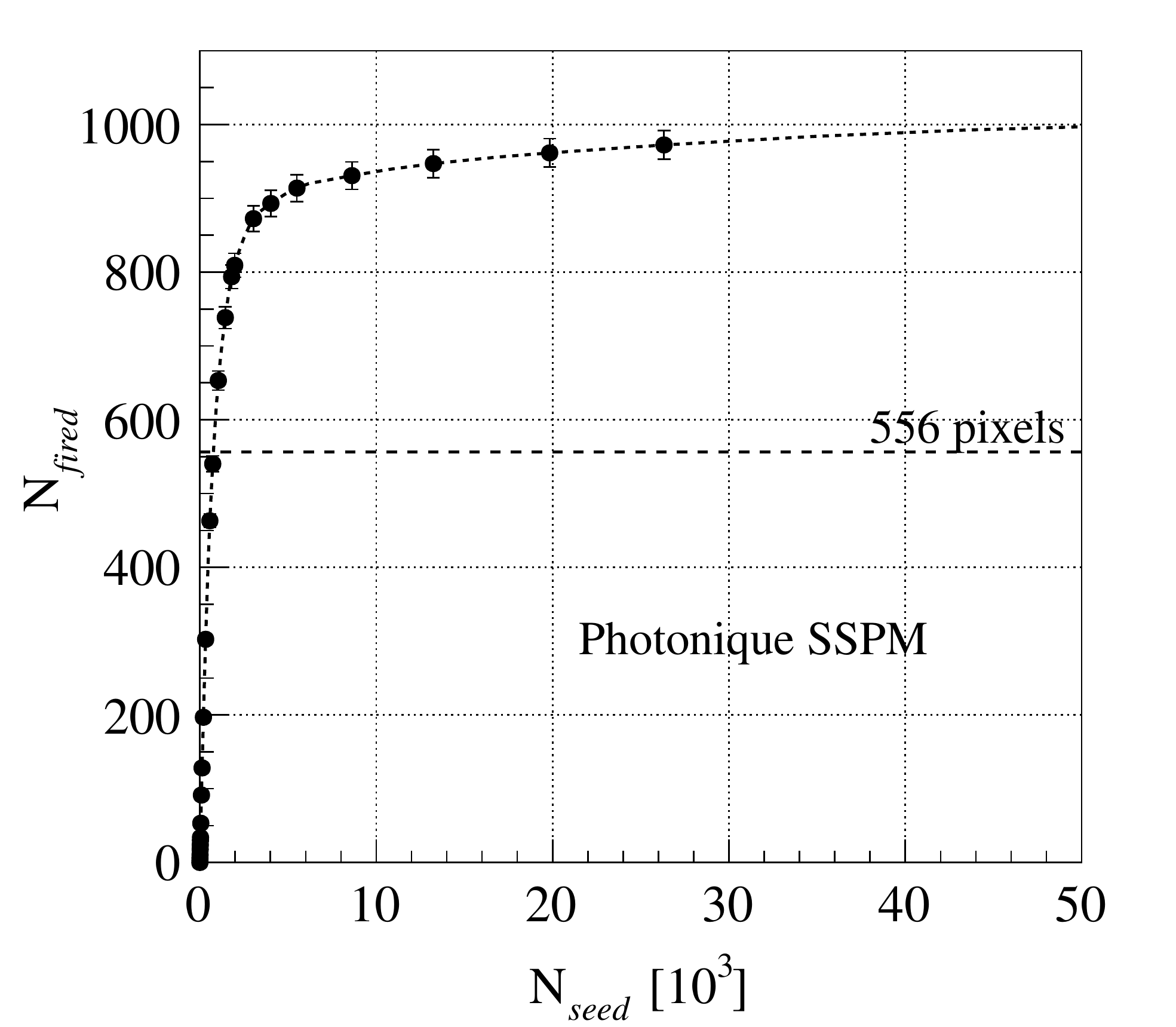}  
  \includegraphics[width=0.34\textwidth]{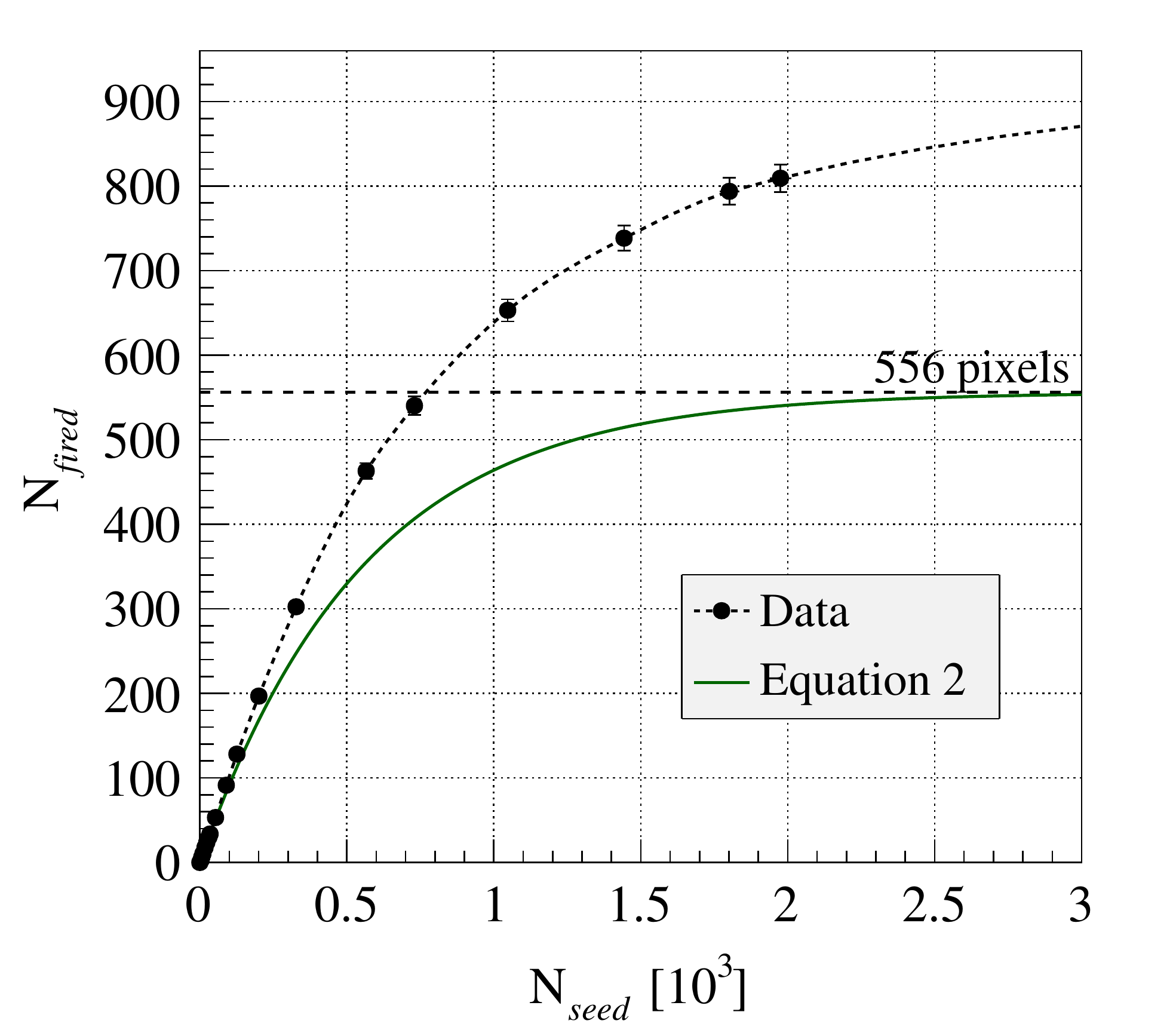}  
  \includegraphics[width=0.34\textwidth]{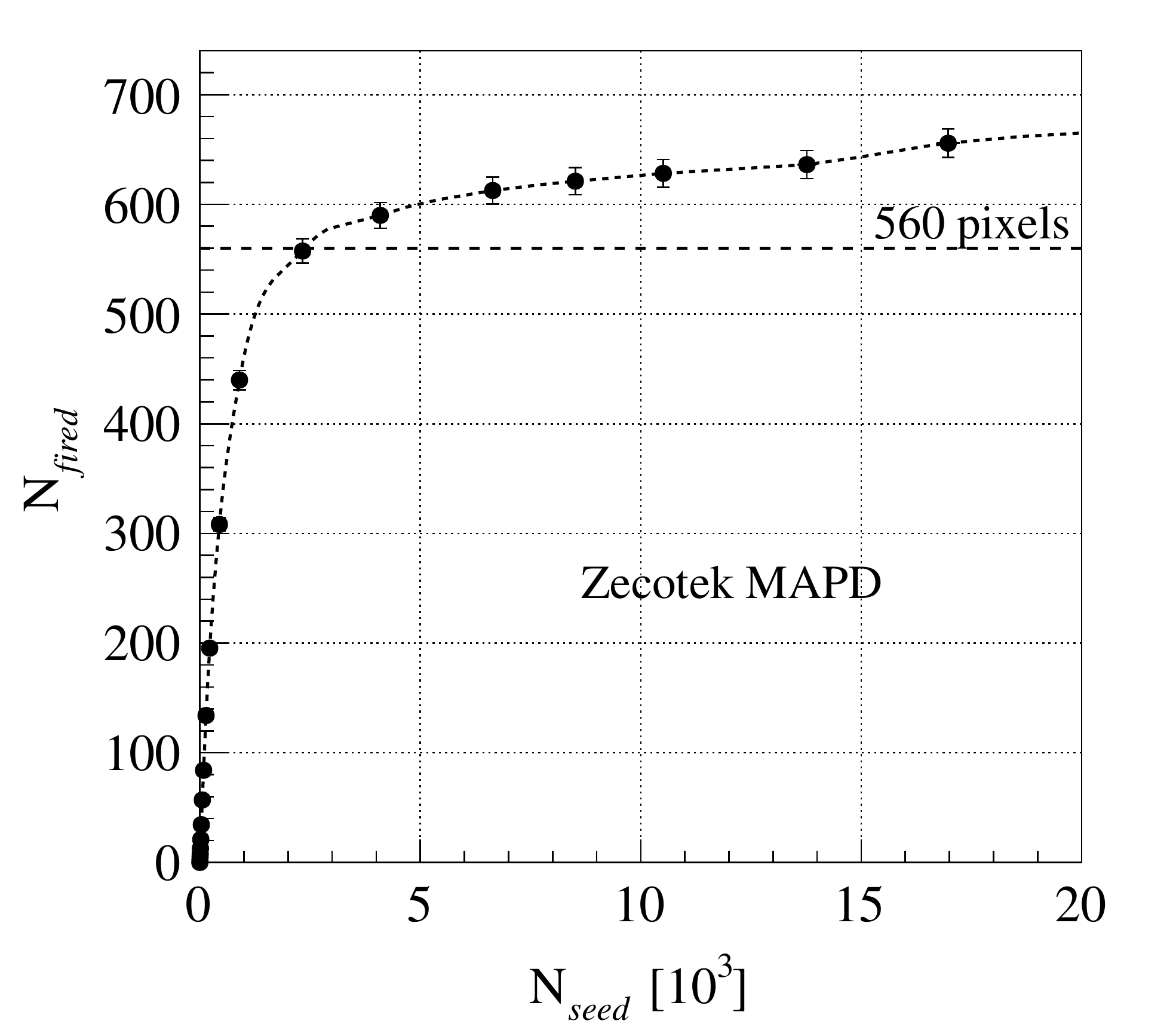}  
  \includegraphics[width=0.34\textwidth]{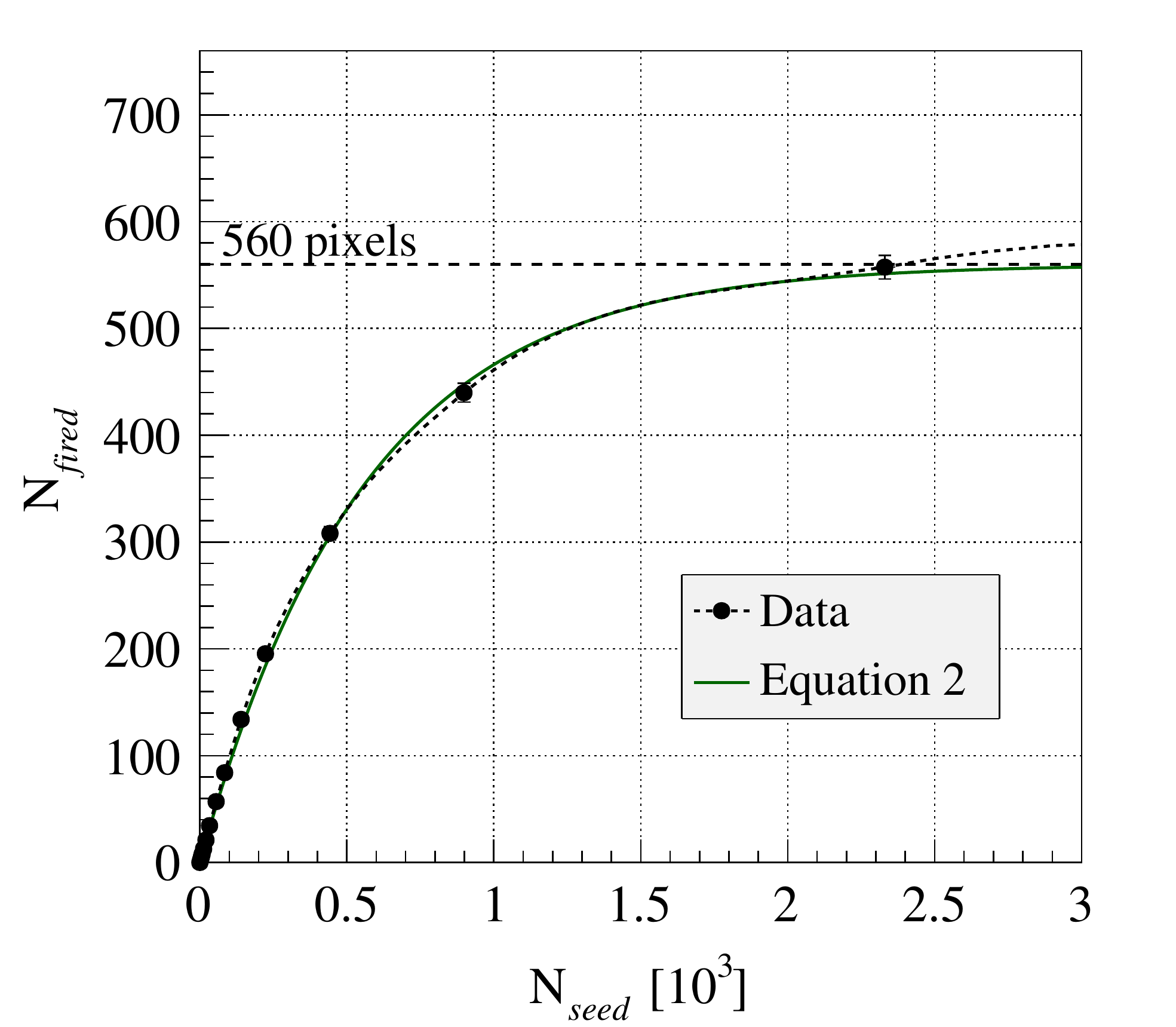}  
  \caption{Response curves of Hamamatsu MPPC with 100 pixels operated at $V_{over} = 0.5\,\mathrm{V}$, Hamamatsu MPPC with 400 pixels operated at $V_{over} = 1.2\,\mathrm{V}$, Photonique SSPM with 556 pixels operated at $V_{over} = 1.2\,\mathrm{V}$ and Zecotek MAPD-1 with 560 pixels operated at $V_{over} = 0.7\,\mathrm{V}$ (top to bottom) for high light intensities (left) and low to medium light intensities (right). The data points are compared with the result of a model calculation given by Eq.~\ref{eq:dynamicrange_noPDE}. The expected values of saturation are indicated by the horizontal dashed lines.}
  \label{fig:dynamicrange_noPDE}
\end{figure*}

Fig.~\ref{fig:dynamicrange_noPDE} shows the measured response curves, $N_{\it fired}$ as a function of $N_{\it seed}$, for the Hamamatsu MPPC with 100 pixels and 400 pixels, as well as for the Photonique SSPM with 556 pixels and Zecotek MAPD with 560 pixels. $N_{\it fired}$ is the direct observable of our setup, while $N_{\it seed}$ is determined via measuring the PIN photodiode output current as described in Section~\ref{sec:method}. However, $N_{\it fired}$ and $N_{\it seed}$ are both derived from the measurement, and therefore both variables include the actual PDE of the respective SiPM. In Fig.~\ref{fig:dynamicrange_noPDE} the same data is presented always with wide range (left) and narrow range (right) of input photons. A spline curve is added to guide the trend of the response curve. The curve expected from the model as in Eq.~\ref{eq:dynamicrange_noPDE} is also drawn in the figure for comparison. The expected outputs approach exponentially the maxima, $N_{\it total}$, which are indicated by the horizontal dotted lines. 

For all sensors we tested, the data agree well with the expected response curve at low photon intensity, however, soon start to diverge. It is clearly visible that the pulse height exceeds the maximum expected value. One notices also that the effect of this over saturation behavior varies among the SiPMs tested here. The maximum output we obtained from Hamamatsu 100U (100 pixels) and Hamamatsu 050U (400 pixels) amounts roughly 200 p.e. (photoelectrons) equivalent and 800 p.e. equivalent, respectively, which would be the expected output for a sensor having twice the number of pixels. On the other hand, the Zecotek device with $N_{\it total} = 560$ seems less affected and the maximum measured output is $\sim$\,650 p.e. equivalent, that amounts $\sim$\,15\,\% larger output than expected. It is also to be noted that within the maximum light intensity ($\sim$\,100\,k "seeds"), we see no clear sign of full saturation of the device, especially in the case of the Hamamatsu 100U. Looking at the response curves of the MPPC with 400 pixels and the SSPM with 556 pixels, one can observe an additional enhancement of the dynamic range compared to the model calculation.

For better comparison, the SiPM response curves are normalized to the total number of pixels corresponding to the respective device. Then the expectation curve appears as Eq.~\ref{eq:dynamicrange_universal}:

\begin{equation}
  \centering
  \frac{N_{\it fired}}{N_{\it total}} = 1 - \exp\left(-\frac{N_{\it seed}}{N_{\it total}}\right)
  \label{eq:dynamicrange_universal}
\end{equation}

\begin{figure*}[t]
  \centering
  \includegraphics[width=0.4\textwidth]{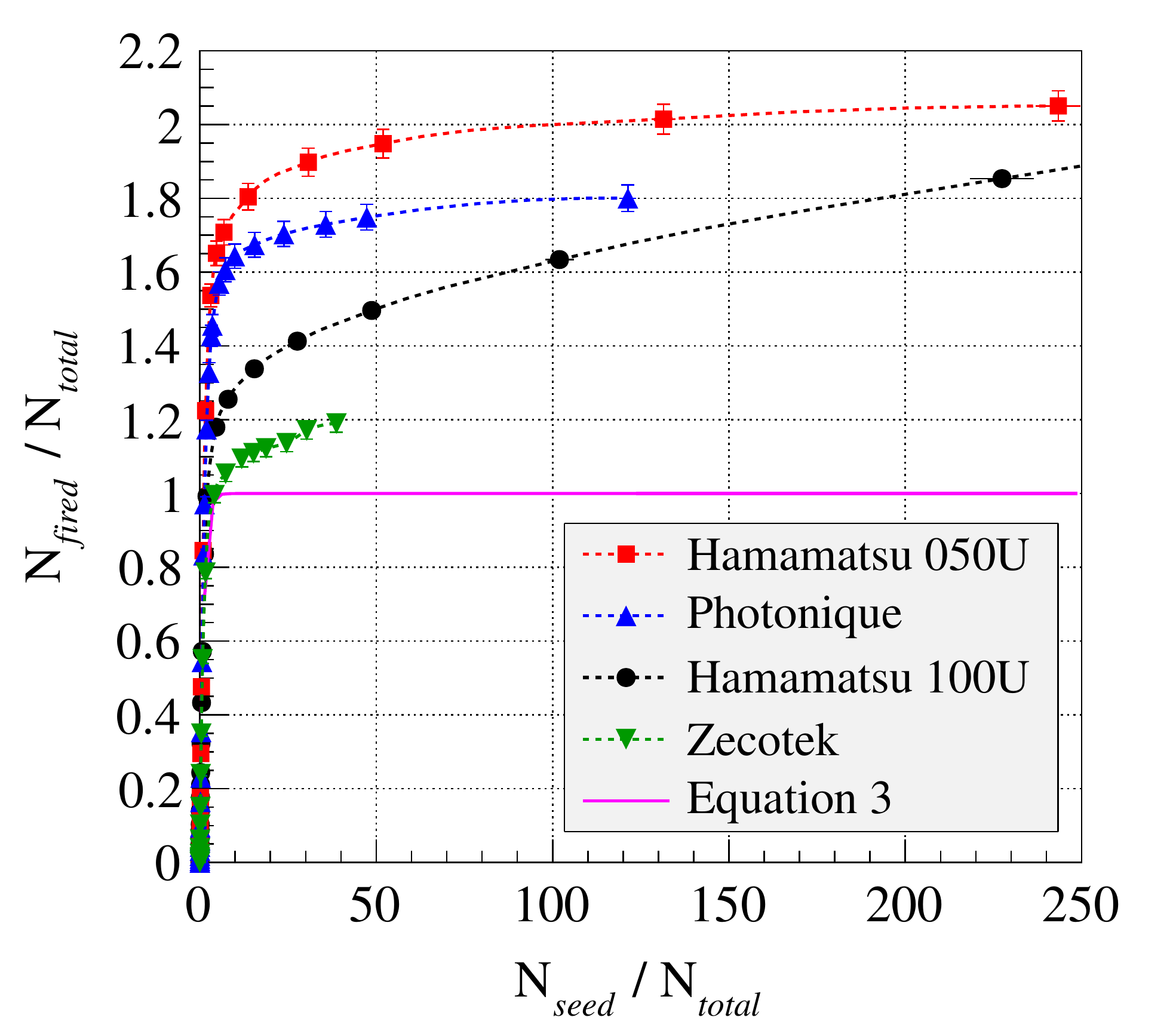}
  \includegraphics[width=0.4\textwidth]{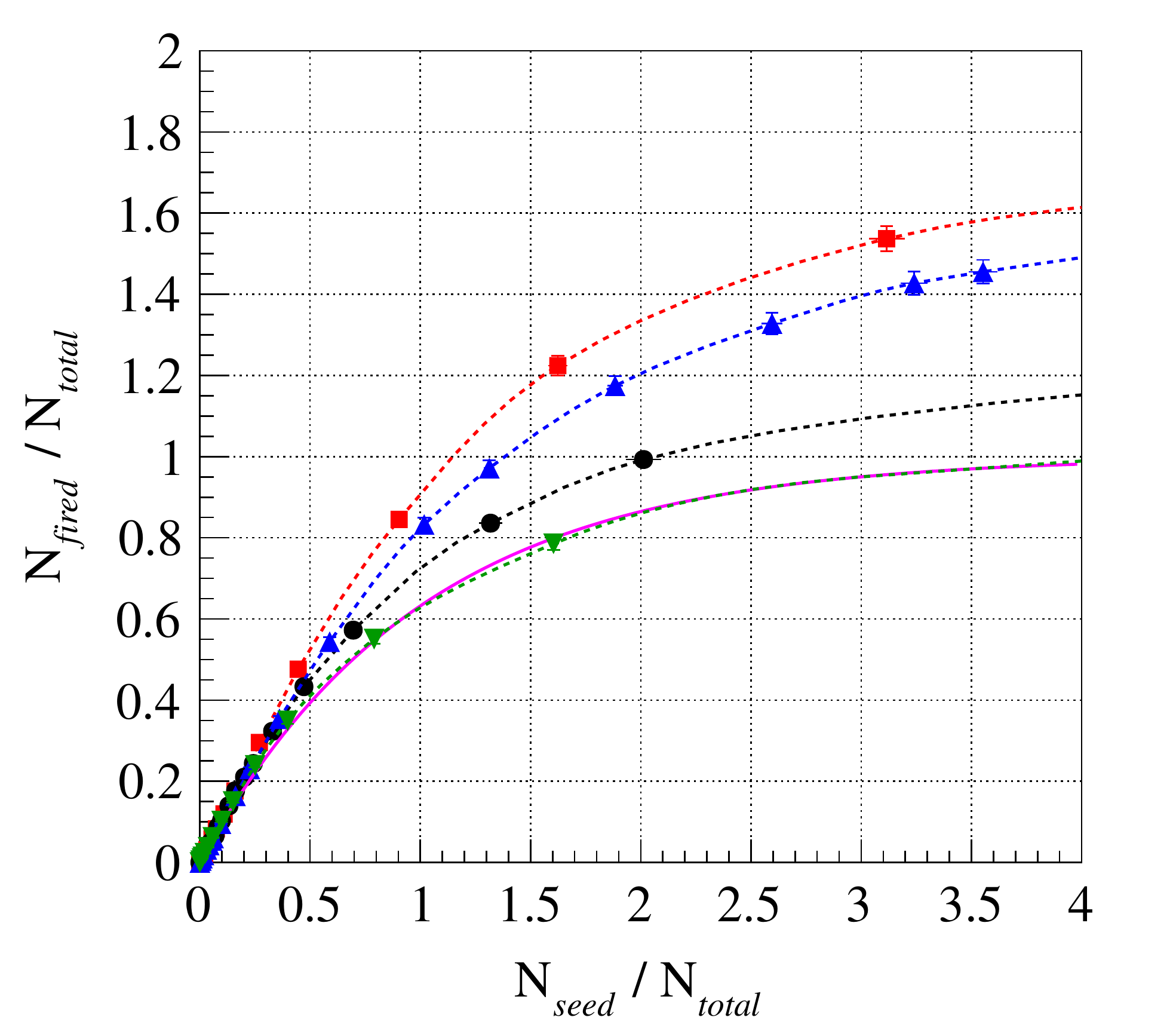}
  \caption{Response curves of various SiPMs, normalized to the total number of pixels of each device, $N_{\it total}$, for high (left) and low to medium light intensities (right).}
  \label{fig:dynamicrange_normPDE}
\end{figure*}

In this representation the response curve is universal to all types of SiPMs. The data from all four SiPMs as well as the universal response function (Eq.~\ref{eq:dynamicrange_universal}) are overlaid and compared in Fig.~\ref{fig:dynamicrange_normPDE}. The plot shows clearly that the degree of the over saturation differs from one SiPM model to the other (Fig.~\ref{fig:dynamicrange_normPDE} left). Within the plotted range, the largest effect is seen in Hamamatsu 050U, followed by Photonique, Hamamatsu 100U and Zecotek. At very high light intensities ($N_{\it seed}/N_{\it total} > 400$), the output of the Hamamatsu 100U even exceeds the one measured for the Hamamatsu 050U device (Fig.~\ref{fig:dynamicrange_noPDE} left). At low light input, the SiPM response is linear and following the model curve, with increasing light intensity the data start to diverge from the expected behavior. The Zecotek sensor appears to be the only device following the expected function at light inputs up to $N_{\it seed}/N_{\it total} = 4$ (Fig.~\ref{fig:dynamicrange_normPDE} right).

\begin{figure*}[t]
  \centering
  \includegraphics[width=0.4\textwidth]{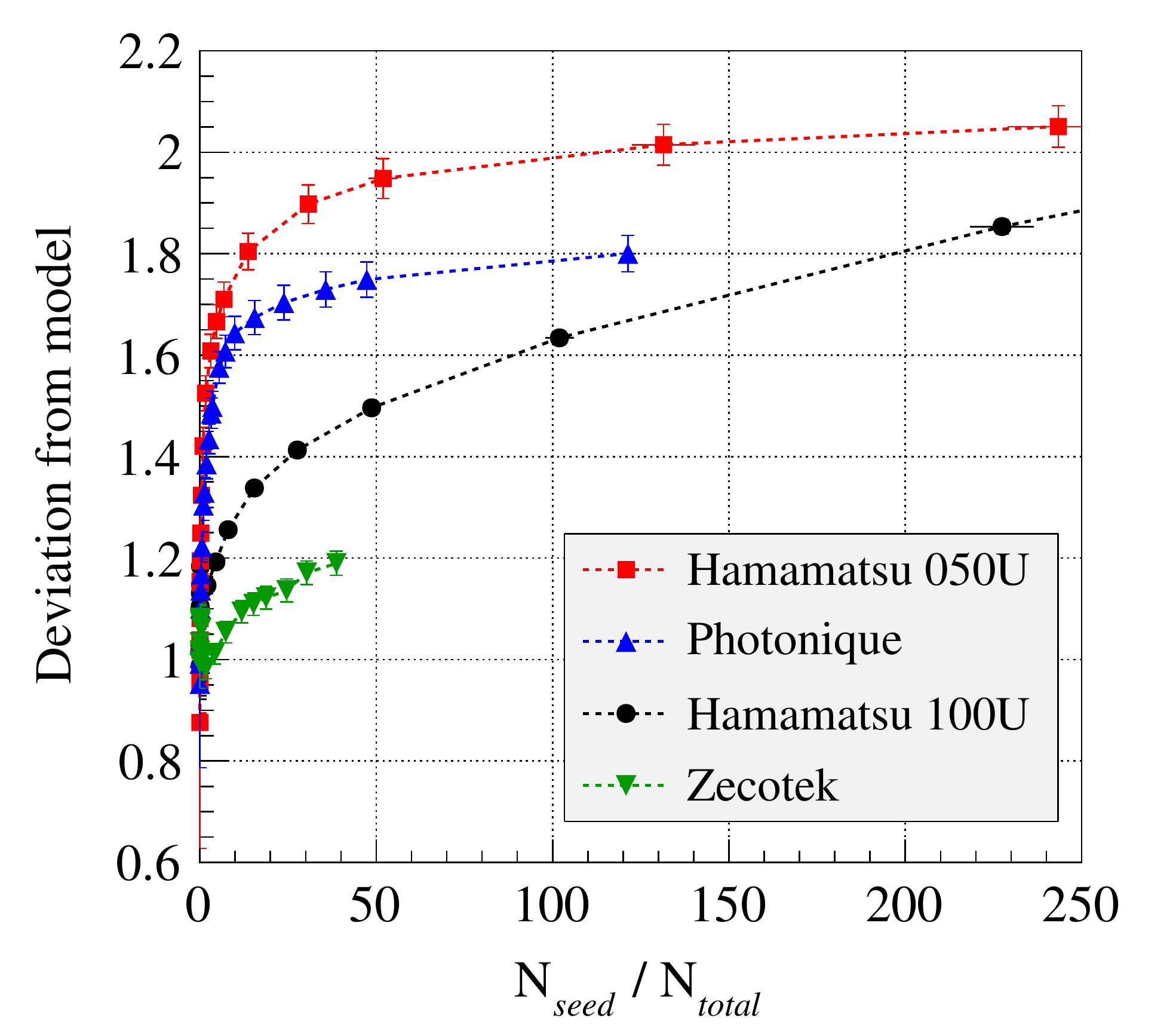}
  \includegraphics[width=0.4\textwidth]{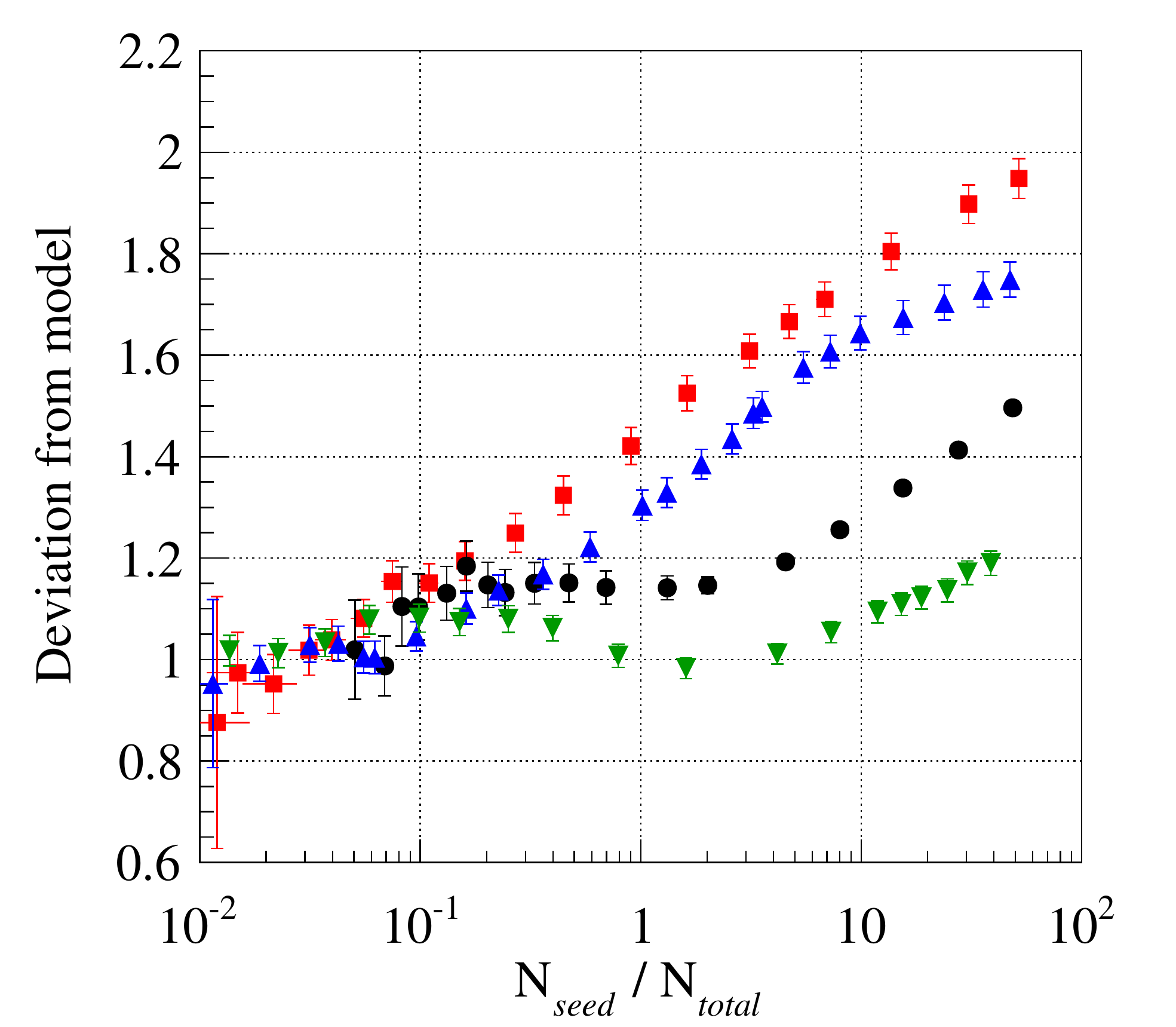}
  \caption{Deviation of the response curves of various SiPMs from the model curve given by Eq.~\ref{eq:dynamicrange_universal}. The results are normalized to the total number of pixels of each device, $N_{\it total}$. The deviation is defined as the ratio between measured and calculated values.}
  \label{fig:dynamicrange_ratio}
\end{figure*}

In order to emphasize the degree of deviation from the theoretical function, we normalized Fig.~\ref{fig:dynamicrange_normPDE} using the function given by Eq.~\ref{eq:dynamicrange_universal}. The results are shown in Fig.~\ref{fig:dynamicrange_ratio}. At low light intensity ($N_{\it seed}/N_{\it total} < 0.3$), where actually the sensors are commonly used, the deviations are very small and all SiPM respond in a similar manner (Fig.~\ref{fig:dynamicrange_ratio} right). In this region it seems that the deviation from the theoretical function increases monotonically for all SiPMs. However, for higher light input we notice two qualitatively different tendencies of deviation. Around $N_{\it seed}/N_{\it total} \sim 0.5$ the deviation starts to decrease and tends to return to the expected value for Hamamatsu 100U and Zecotek sensors, before increasing again. For the other two SiPMs the deviation increases monotonically. It is also to be noted that even the two Hamamatsu MPPCs, which are supposed to have a comparable response, show a different behavior. 

\begin{figure*}[t]
  \centering
  \includegraphics[width=0.4\textwidth]{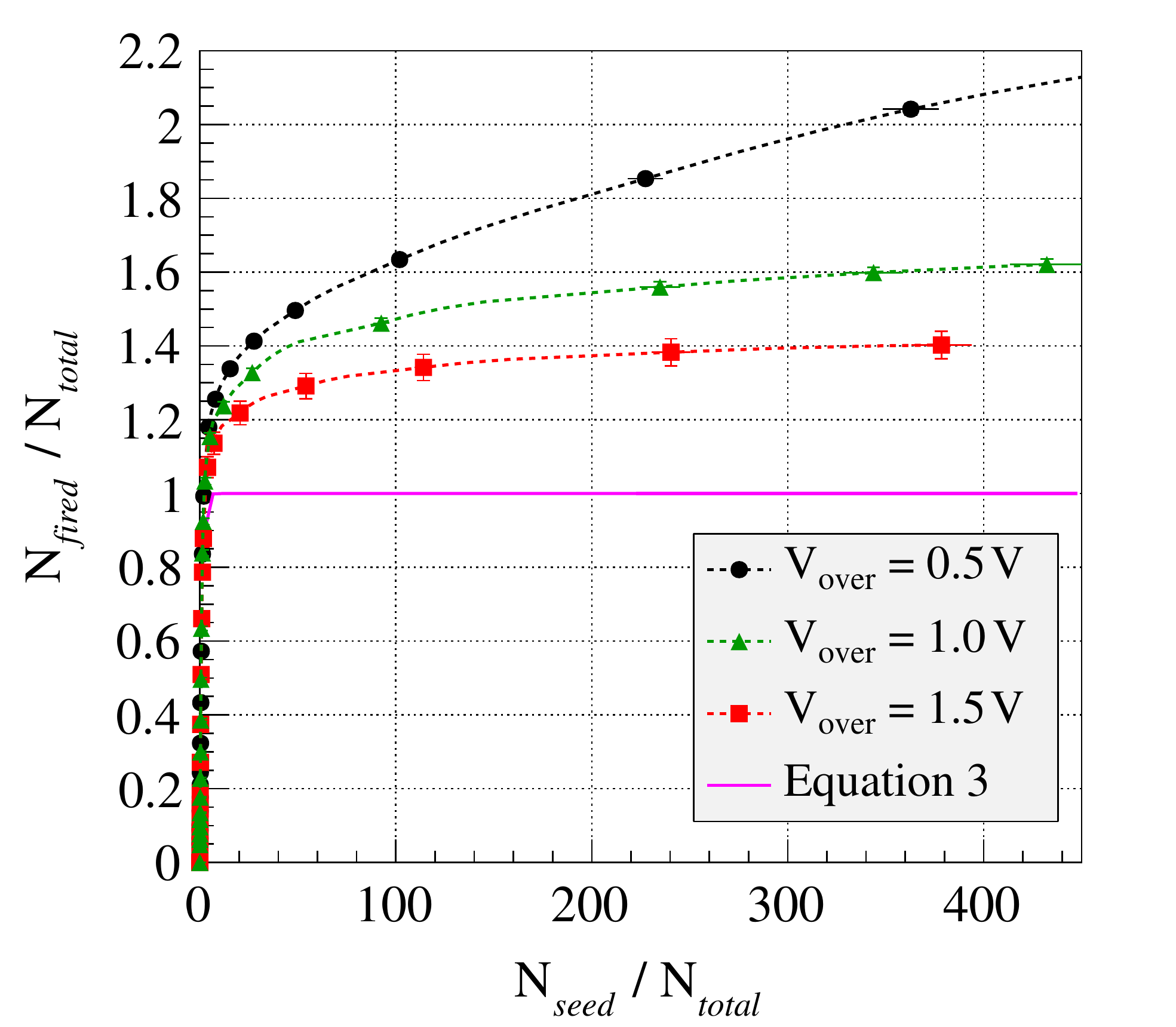}
  \includegraphics[width=0.4\textwidth]{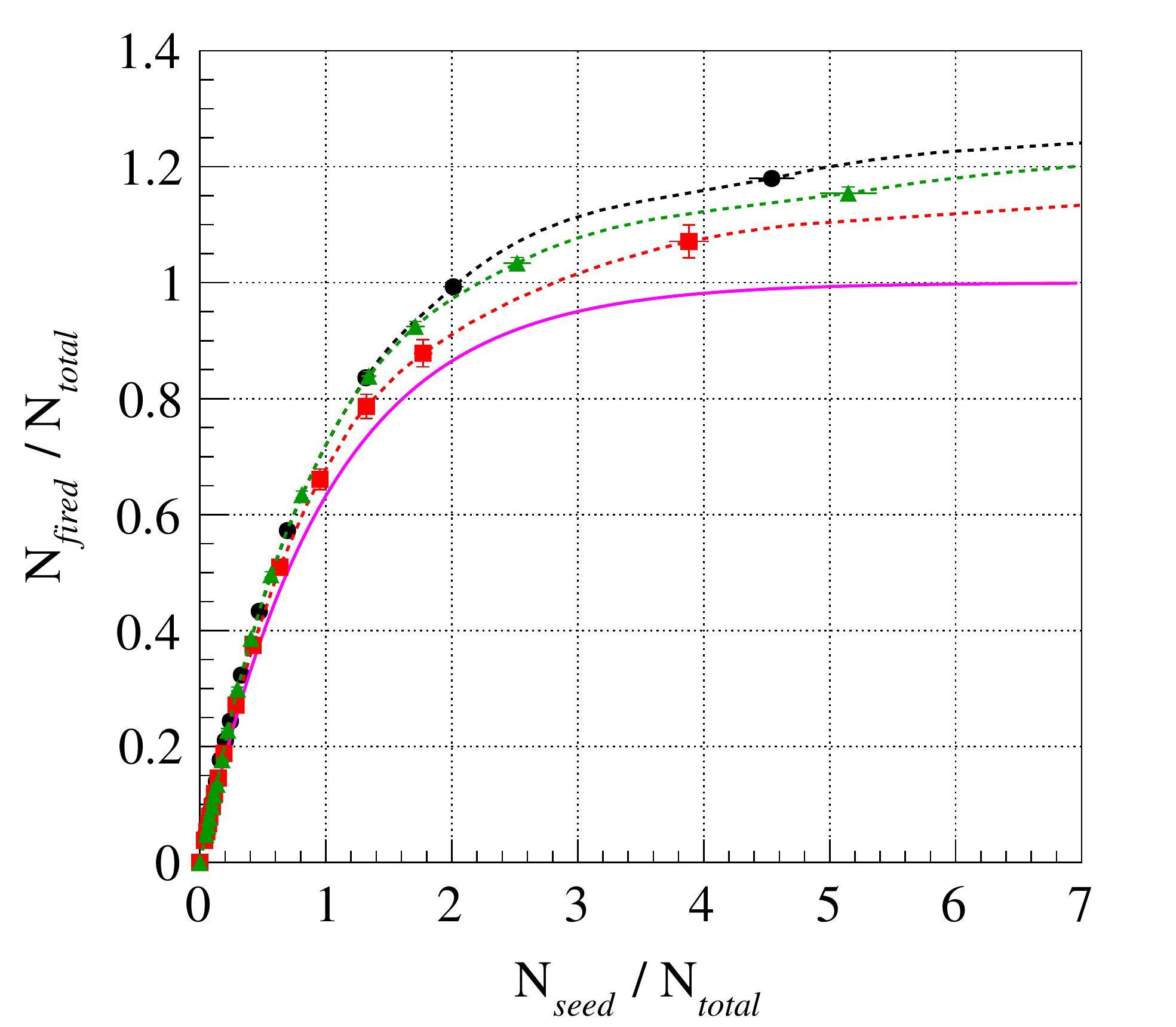}
  \caption{Over-voltage dependency of the response curve for Hamamatsu 100U. The data are compared with the model calculation given by Eq.~\ref{eq:dynamicrange_universal}. The results are normalized to the total number of pixels of the device, $N_{\it total}$.}
  \label{fig:dynamicrange_voltage}
\end{figure*}

In a last step, we determined the over-voltage dependency of the SiPM response, as shown in Fig.~\ref{fig:dynamicrange_voltage} for the Hamamatsu 100U. At low light intensities ($N_{\it seed}/N_{\it total} < 0.3$) all curves are following the model given by Eq.~\ref{eq:dynamicrange_universal} (Fig.~\ref{fig:dynamicrange_voltage} right). Increasing the light intensity the response curves deviate from the model calculation. Moreover, the deviation is strongly correlated with the applied over-voltage, especially for very high light intensities (Fig.~\ref{fig:dynamicrange_voltage} left).

\section{Discussion}
\label{sec:discussion}
In Section~\ref{sec:results}, the raw data of the measurements ($N_{\it fired}$ and $N_{\it seed}$) are used to describe the SiPM response curves. By using the PDE of the respective device, we can obtain an alternative representation of the response curves, i.e. we can plot $N_{\it fired}$ as a function of $N_{\it photon}$, the number of incident photons. As illustrated in Fig.~\ref{fig:method}, the relation between $N_{\it seed}$ and $N_{\it photon}$ is given by $N_{\it photon} = N_{\it seed}/\mathrm{PDE}$.

The PDE values for the Hamamatsu MPPCs (72\,\% for 100U, 47\,\% for 050U at 400\,nm) were taken from the data sheet~\cite{hamamatsu}. As here the PDEs are evaluated by a current measurement, which cannot distinguish after-pulsing and cross-talk effects, the values are overestimated. These two effects would amount in an overestimation of $\sim$\,20\,\% and $\sim$\,10\,\% of the PDE~\cite{discussion}, respectively. For the Photonique SSPM and Zecotek MAPD, PDE values of 18\,\%~\cite{photonique} and 15\,\%~\cite{orth} for a wavelength of 400\,nm were found. There are several other measurements of the PDE available, but the results are also known to depend on the operating conditions, e.g. over-voltage and temperature. Therefore, in order to handle this problem the best, we refer to the PDE values provided by the company and, in case of the Hamamatsu MPPCs, we take those values as an upper boundary of an uncertainty band going down to a $\sim$\,30\,\% smaller value. It's important to stress, that the choice of the PDE value does not affect the measurement result, the PDE is only needed to evaluate $N_{\it photon}$. $N_{\it fired}$ and $N_{\it seed}$ include the actual PDE already, because they are measured quantities.

\begin{figure*}[t]
  \centering
  \includegraphics[width=0.34\textwidth]{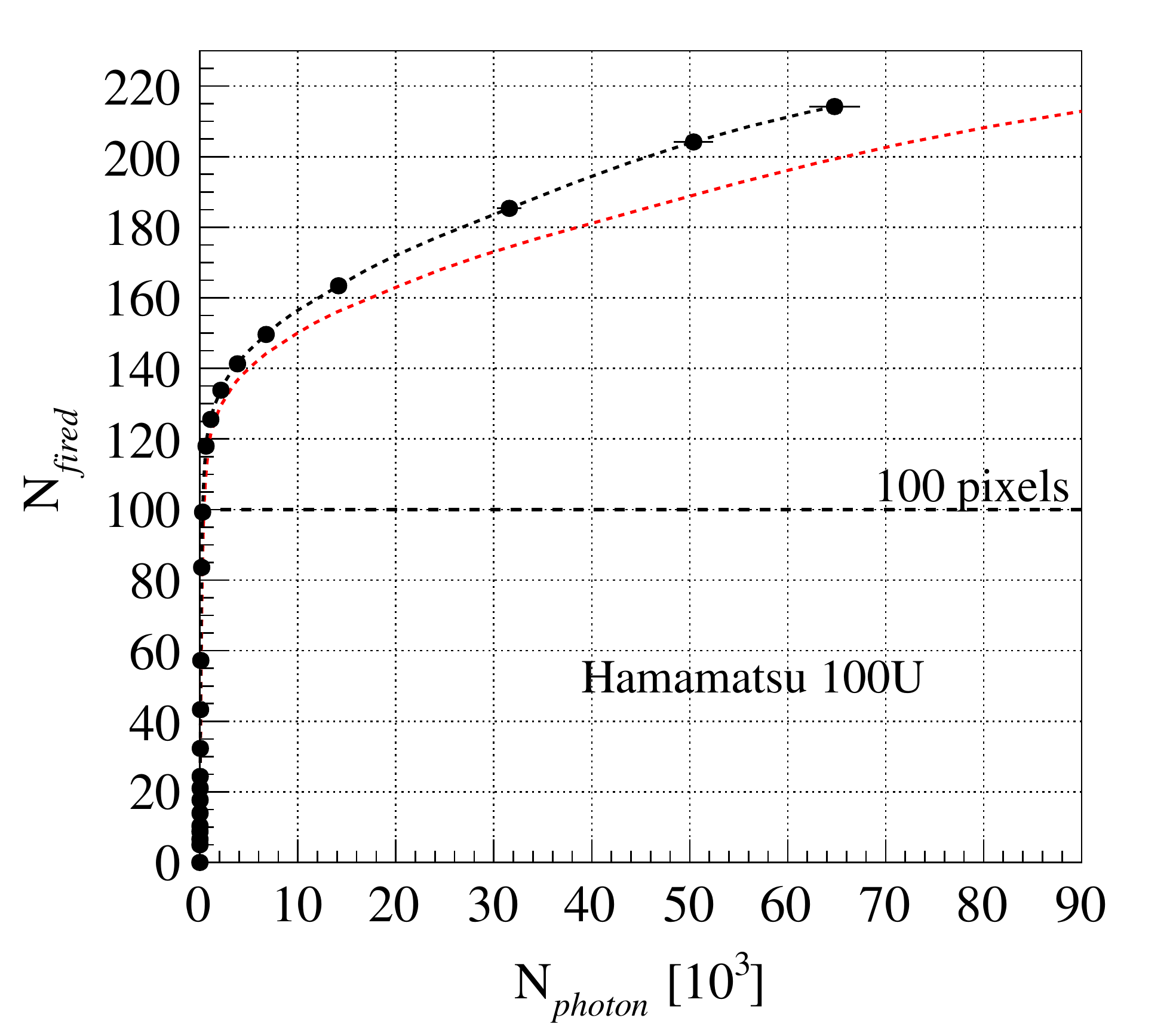}
  \includegraphics[width=0.34\textwidth]{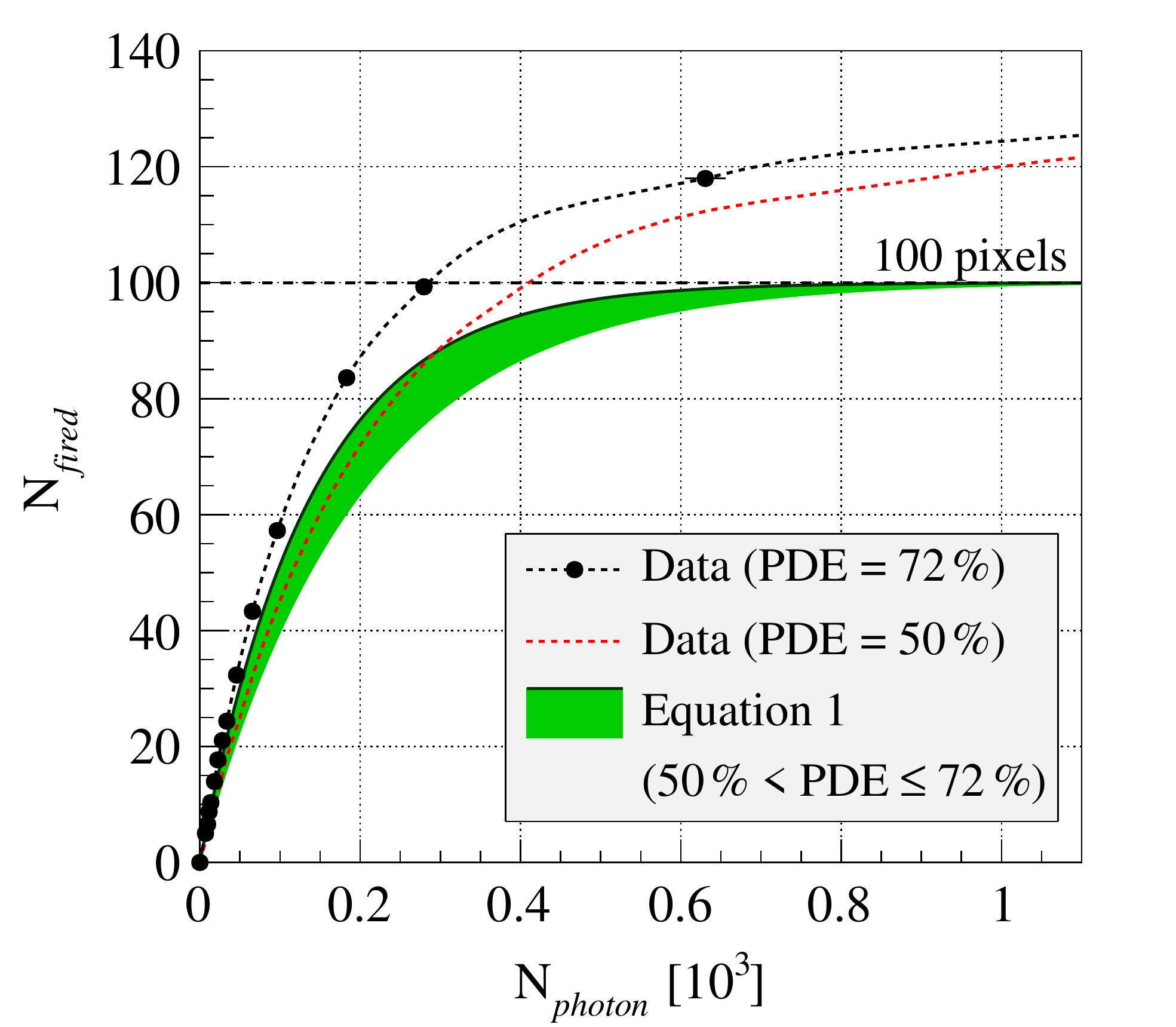}
  \includegraphics[width=0.34\textwidth]{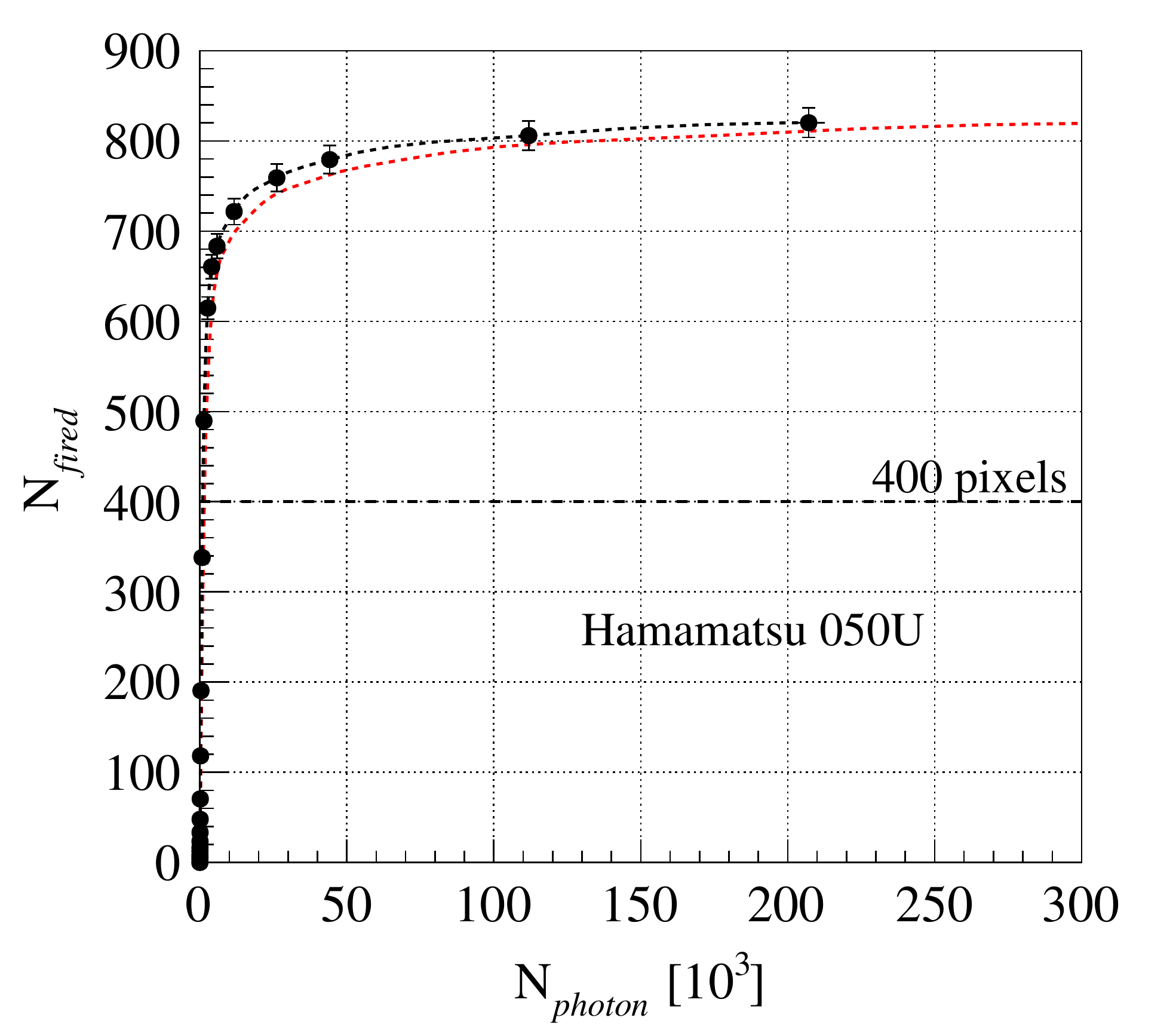}  
  \includegraphics[width=0.34\textwidth]{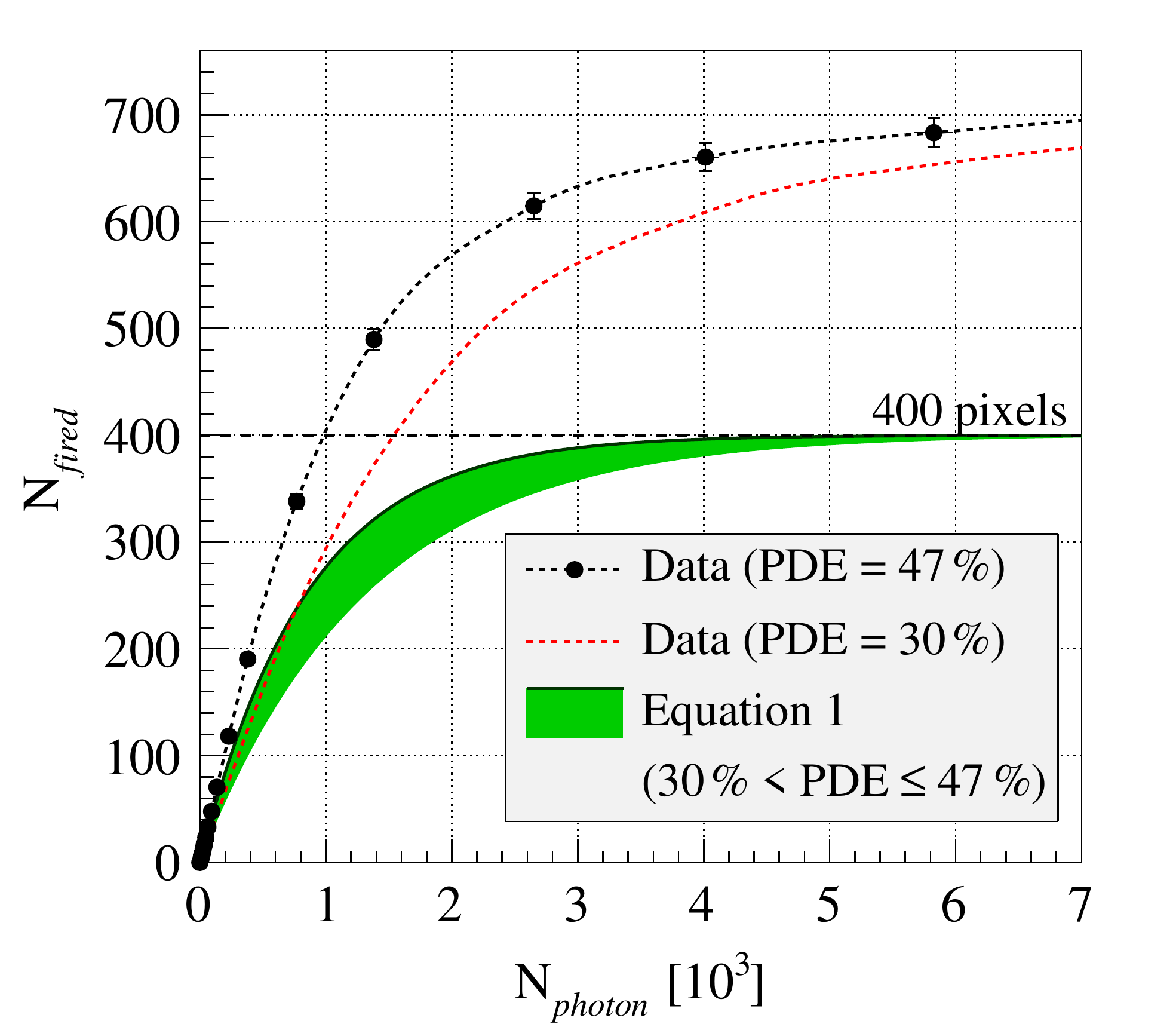} 
  \includegraphics[width=0.34\textwidth]{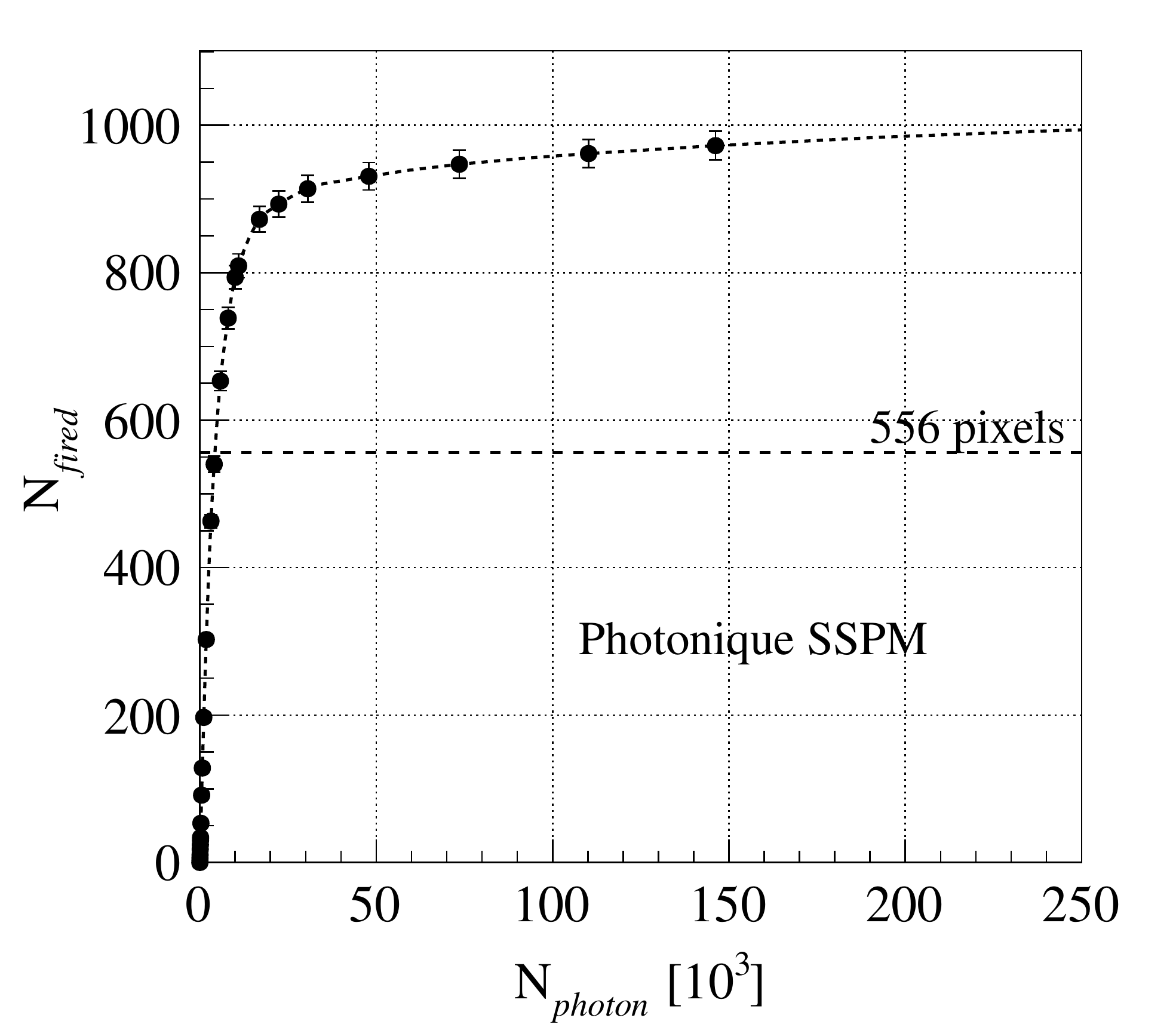}  
  \includegraphics[width=0.34\textwidth]{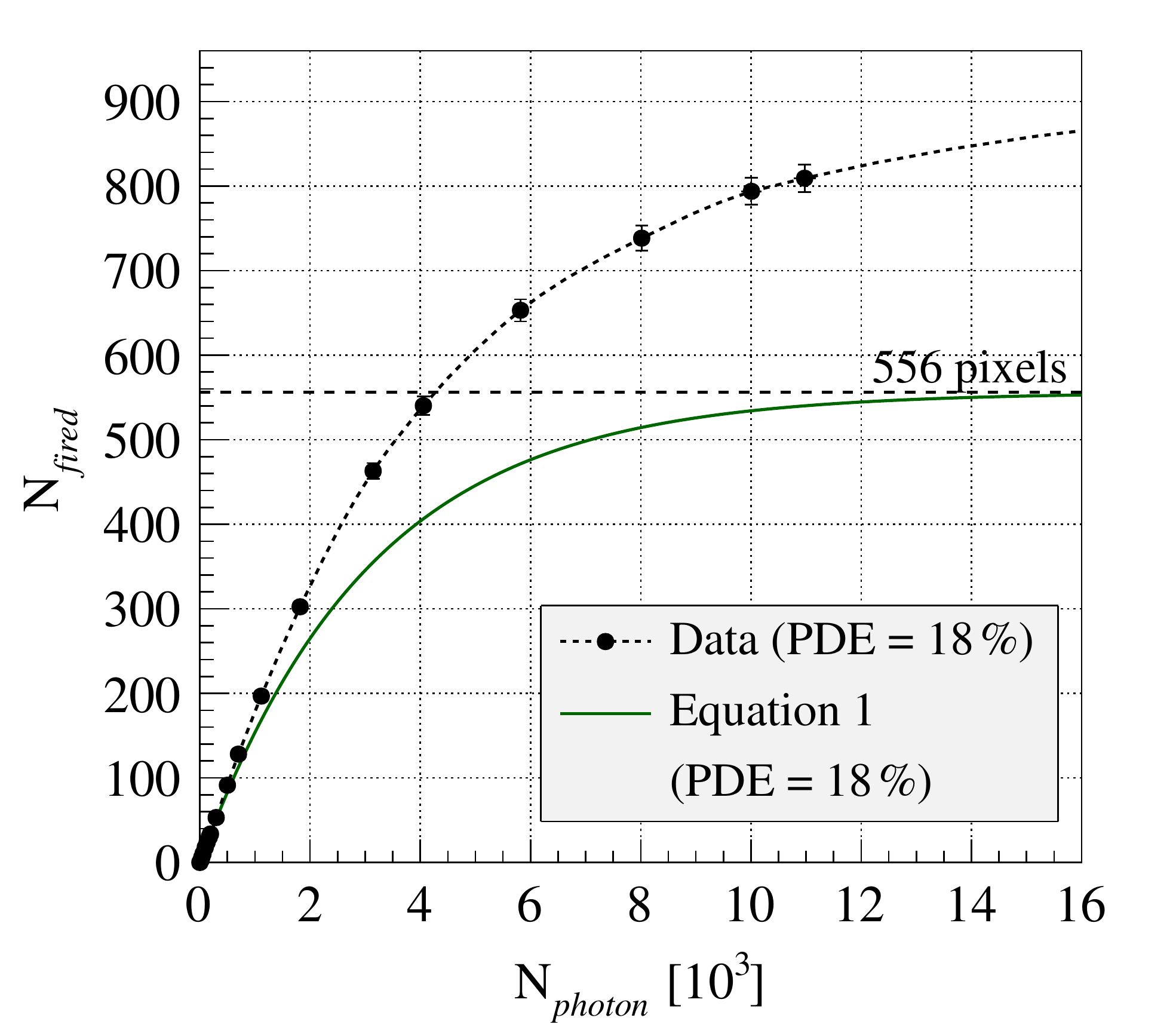}  
  \includegraphics[width=0.34\textwidth]{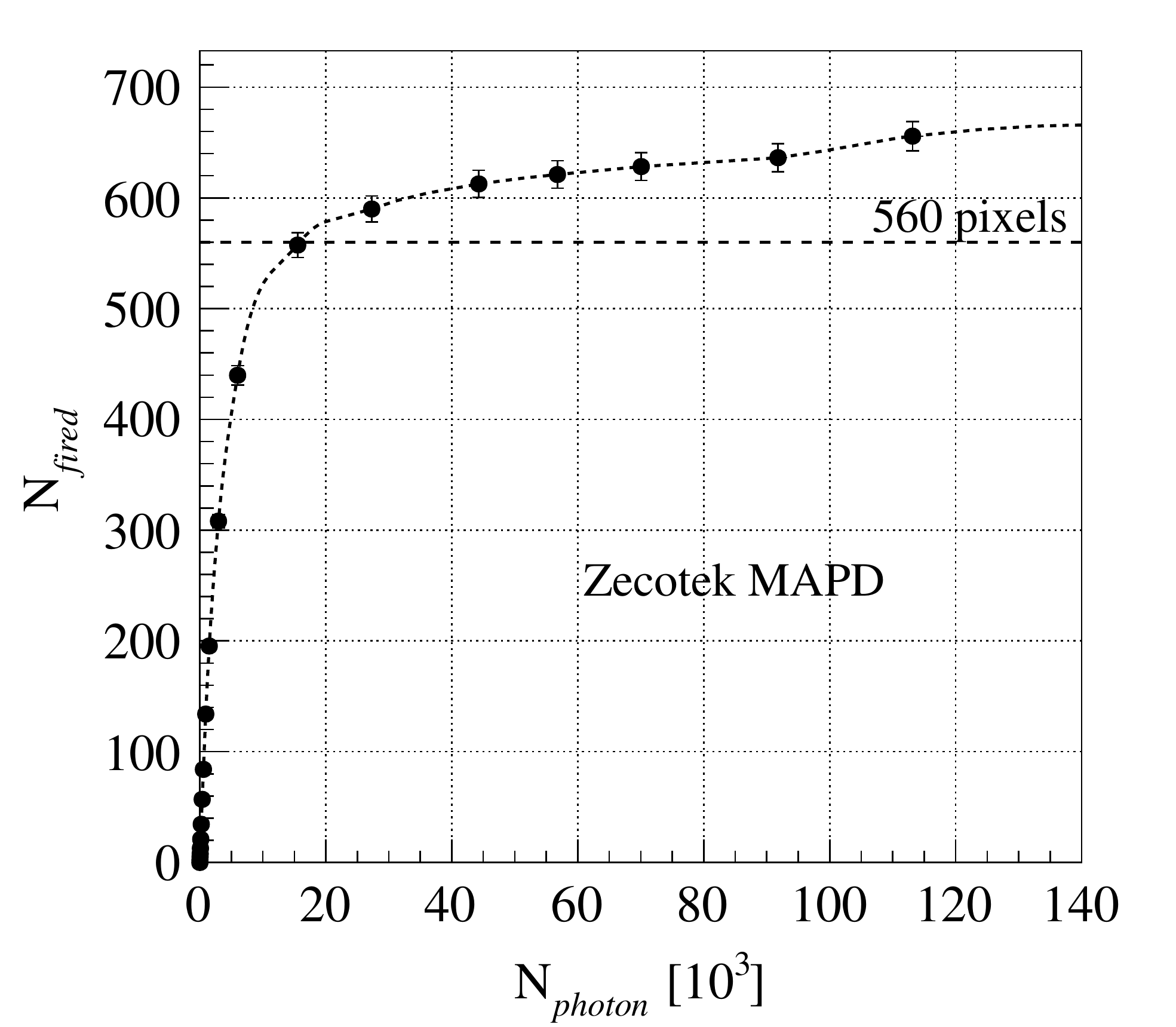}  
  \includegraphics[width=0.34\textwidth]{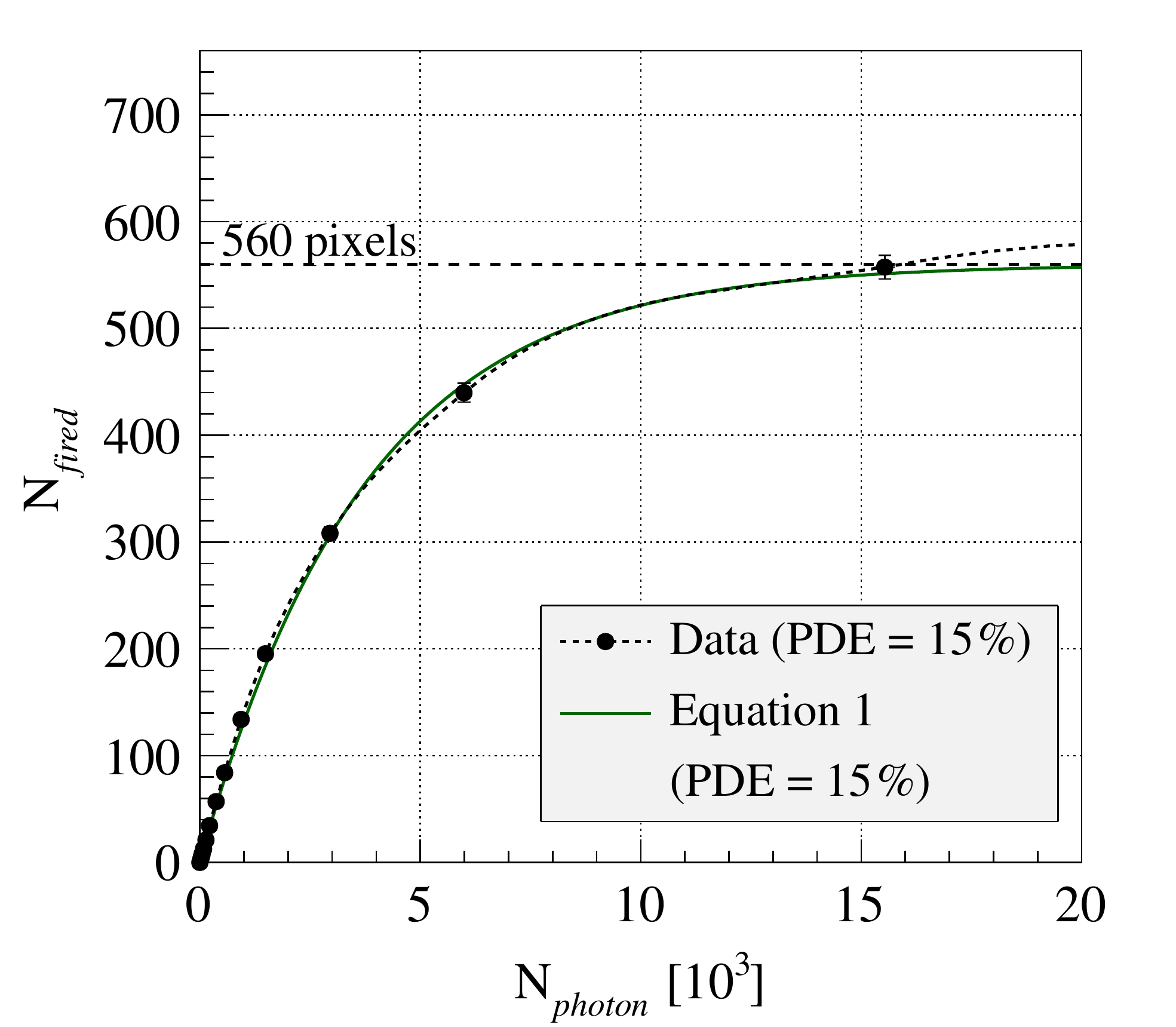}  
  \caption{Alternative representation of response curves ($N_{\it fired}$ as a function of $N_{\it photon}$) for Hamamatsu MPPC with 100 pixels, Hamamatsu MPPC with 400 pixels, Photonique SSPM with 556 pixels and Zecotek MAPD-1 with 560 pixels (top to bottom) for high light intensities (left) and low to medium light intensities (right). The PDE values used to evaluate $N_{\it photon}$ are indicated in the plots. The data points are compared with the result of a model calculation given by Eq.~\ref{eq:dynamicrange}. The expected values of saturation are indicated by the horizontal dashed lines.}
  \label{fig:dynamicrange}
\end{figure*}

Fig.~\ref{fig:dynamicrange} shows the response curves represented as $N_{\it fired}$ as a function of $N_{\it photon}$, for all tested SiPMs. The same data is again presented always with wide range (left) and narrow range (right) of input photons. A spline curve is added to guide the trend of the response curve. The curve expected from the model as in Eq.~\ref{eq:dynamicrange} is also drawn in the figure for comparison. The expected maxima, $N_{\it total}$, are indicated by the horizontal dotted lines. For the plots, $N_{\it photon}$ is calculated by using a certain PDE value. In case of the Hamamatsu SiPMs the data are plotted for two different values of the PDE and the model is calculated within a PDE range, as described above. The slopes of the response and model curves change accordingly. However, the maximum measured output ($N_{\it fired}$) is unaffected by the choice of the PDE value (Fig.~\ref{fig:dynamicrange} left).

\begin{figure*}[t]
  \centering
  \includegraphics[width=0.4\textwidth]{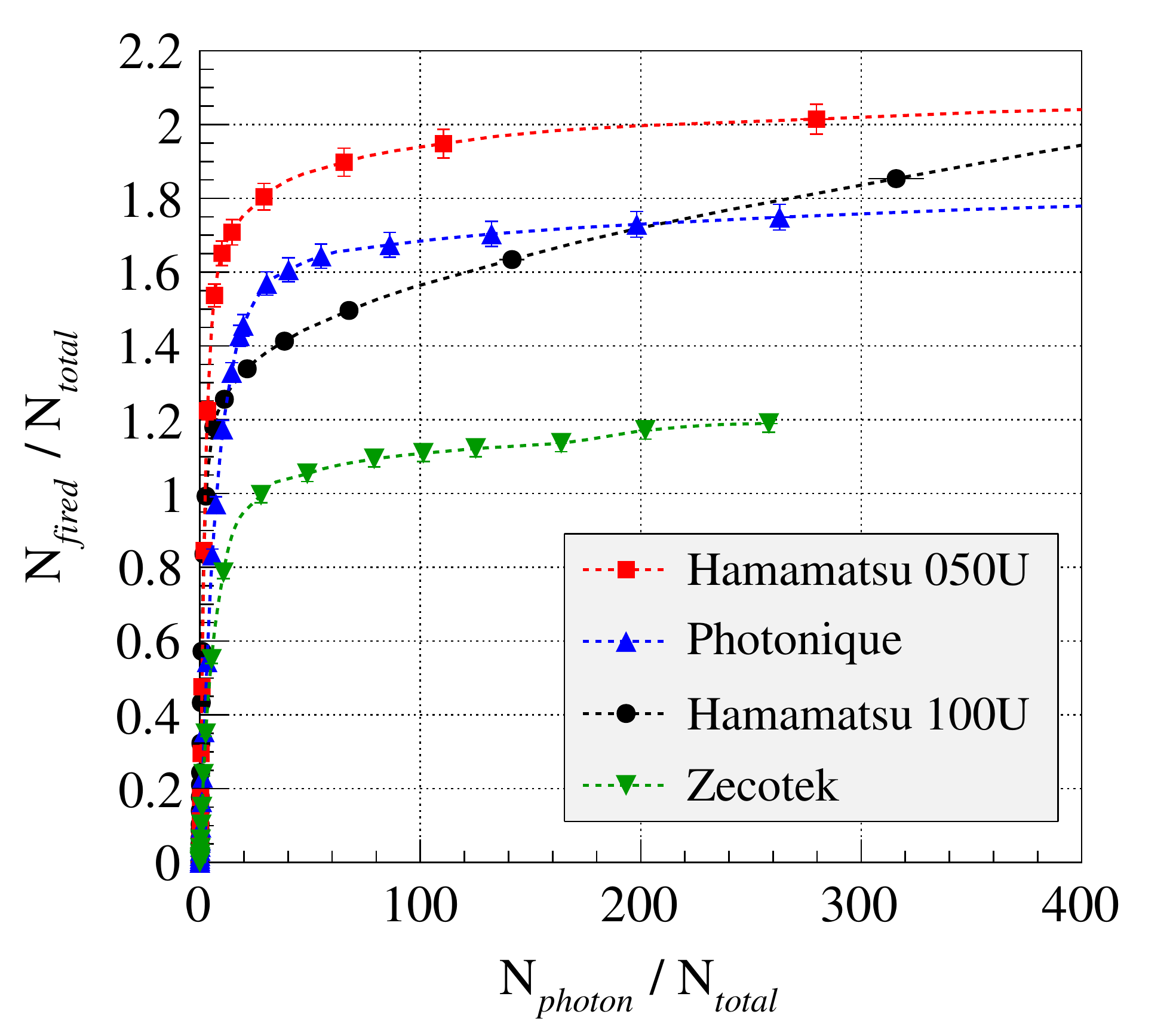}
  \includegraphics[width=0.4\textwidth]{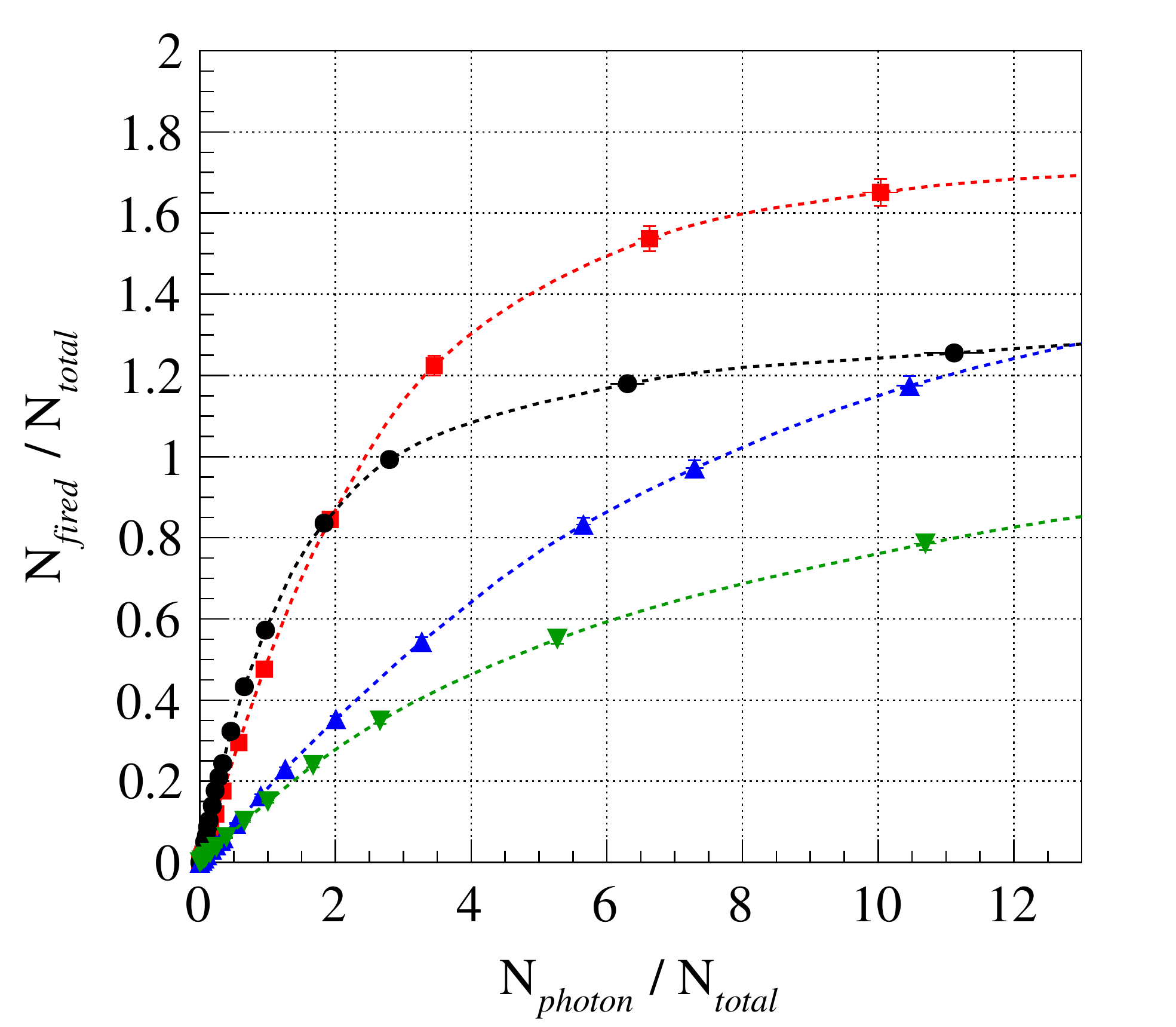}
  \caption{Response curves, represented as $N_{\it fired}$ as a function of $N_{\it photon}$, for various SiPMs, normalized to the total number of pixels of each device, $N_{\it total}$, for high (left) and low to medium light intensities (right).}
  \label{fig:dynamicrange_norm}
\end{figure*}

The response curves of all four SiPMs are again normalized to the total number of pixels corresponding to the respective device and are overlaid in Fig.~\ref{fig:dynamicrange_norm}. The response curves are crossing (Fig.~\ref{fig:dynamicrange_norm} right) because of the different PDE values utilized to evaluate $N_{\it photon}$ on one hand, and the varying strength of the over saturation effect on the other hand. At low number of induced photons, the SiPM response is still linear and the slope of the curve is mainly determined by the PDE used. With increasing light intensity the data start to diverge from the expected behavior (depending on the strength of the effect) and the curves cross. 

Up to now, no convincing explanation for this over saturation and enhanced dynamic range could be found. Other groups report a similar observation. In Ref.~\cite{eckertsimu}, a SiPM response exceeding the total number of pixels at high operating voltage is reported. In this case, however, the number of fired pixels is estimated by integrating the signal using a charge ADC and the result could be explained by the correlated noise (after-pulsing, cross-talk and dark-noise). As discussed, in our case the influence of delayed correlated noise is small, as we are operating the SiPMs at low gain and evaluate the number of pixels from the signal pulse height and not from the collected charge. Cross-talk that occurs almost instantaneously to the initial laser pulse is suppressed for most of the measurements due to high pixel occupancy at high light intensities. Very fast after-pulses occurring within the signal rise time of $\sim$\,2\,-\,3\,ns, may lead to an overestimation of the SiPM output ($N_{\it fired}$) by $\sim$\,1\,-\,2\,\%, and thus cannot explain the large over-saturation we observed. Also the over-voltage dependency of the SiPM response (Fig.~\ref{fig:dynamicrange_voltage}) shows that the measurement is hardly affected by correlated noise: If the measured response was influenced by any of these effects, the over saturation effect would be stronger at higher over-voltage, since the probabilities for after-pulsing, cross-talk as well as for dark-noise would increase. As one can see from Fig.~\ref{fig:dynamicrange_voltage}, this is not the case. 

As described in Ref.~\cite{uozumi}, the SiPM output is strongly depending on the pulse width of the light pulses used to illuminate the sensor. When the pulse width is comparable to or exceeds the pixel recovery time, an enhancement of the dynamic range and an output beyond the total number of pixels can be observed. Since we are using a pulse width of 32\,ps (FWHM), we are not influenced by the recovery process. 

To exclude effects from electronics, in particular a non-linear behavior of the preamplifier at large input signals, the linearity of the preamplifier was confirmed in a measurement, which resulted in a mean gain of $23 \pm 0.8$ for the whole input range we tested. It is also to be noted that, if an input signal to the preamplifier exceeded the linear region, a pulse height measurement would underestimate the actual signal because of the saturation of the preamplifier output, which means that one would even underestimate the real over saturation. Eventually, the over saturation behavior was observed when the SiPM signal was not amplified but directly fed into the oscilloscope. 
 
One possible explanation could be that a very high number of input photons per pixel may trigger several avalanches simultaneously, giving rise to a slightly higher output signal compared to the single photon signal. However, the fact that even the two MPPCs do not show the same behavior is in contrast to this assumption and indicates a more complex effect behind.

Another possible reason for the observed effect might be related to the region in-between the microcells. The trenches separating the individual pixels are coated with a thin reflective layer of aluminum and are supposed to be insensitive to incoming light. However, at very high light exposure, some photons may pass the layer, resulting in an additional signal. In order to explain the observed over saturation solely by this effect, one has to assume a gain of at least 10$^{3}$ for these inactive regions, which is questionable because of the low field there. However, such an effect could explain the discrepancy in the behavior of the two MPPCs. The Hamamatsu 050U has a lower fill factor, i.e. a larger inactive area, than the Hamamatsu 100U.

In general one should note that SiPMs are typically not operated in the regime of very high light exposure since the output linearity is lost, but are preferably used to measure low amounts of light, in the linear, dynamic range, where measurement and model agree well. Therefore, most cases of application would not be affected by our observation. Nevertheless, understanding this behavior is going to advance the overall understanding of this still relatively new device, and may open a new application, e.g. it may allow to use SiPMs for a wide range of light intensities, using calibration curves, of course with the drawback of decreasing accuracy for increasing intensity.  

\section{Conclusion}
\label{sec:conclusion}
Our results show the SiPM signal response for a wide range of light intensity. For low light levels, the dynamic range of the photons follow the expected behavior as given by the model equation and simulation. With increasing light intensities the signal response starts to diverge from the predicted values and exceeds the expected maximum by a factor between 1 and 2. Furthermore, at light intensities reaching 500 times the number of pixels, still no full saturation was observed. This behavior was found for all tested devices but varies in magnitude. It suggests that the current understanding of SiPMs, i.e. each pixel acts as a digital device giving 0 or 1 output regardless of the number of induced photons onto the pixel, might be rather naive and we have to improve our understanding of the device.

\section*{Acknowledgements}
This work is partly supported by the EU Projects HadronPhysics2 (project 227431) and HadronPhysics3 (project 283286).

\end{document}